\documentclass[twocolumn,aps,reprint,showpacs]{revtex4}

\usepackage{graphicx}
\usepackage{dcolumn}
\usepackage{bm}
\usepackage{amsmath}
\usepackage{subfig}
\usepackage{color}
\usepackage{amssymb}
\usepackage{txfonts}
\usepackage{wasysym}
\usepackage{caption}
\usepackage[font={small}]{caption}

\definecolor{antiquebrass}{rgb}{0.8, 0.58, 0.46}
\definecolor{orange}{rgb}{1, 0.5, 0}
\definecolor{halayaube}{rgb}{0.4,0.22,0.33}
\definecolor{officegreen}{rgb}{0.0,0.5,0.0}

\begin{document}

\title{Glassy dynamics in asymmetric binary mixtures of hard-spheres}

\author{Edilio L\'azaro-L\'azaro$^{1}$, Jorge Adri\'an 
Perera-Burgos$^{2}$, Patrick Laermann$^{3}$, Tatjana Sentjabrskaja$^{3}$,  
Gabriel P\'erez-\'Angel $^{4}$, Marco Laurati$^{3,5}$, Stefan U. 
Egelhaaf$^{3}$, Magdaleno Medina-Noyola$^{1,5}$, Thomas 
Voigtmann$^{6,7}$, Ram\'on Casta\~neda-Priego $^{5,a}$ and Luis Fernando 
Elizondo-Aguilera$^{7,b}$} 
\affiliation{$^{1}$ Instituto de F\'isica Manuel Sandoval Vallarta,
Universidad Aut\'onoma de San Luis Potos\'i, Alvaro Obreg\'on 64,
78000, San Luis Potos\'i, San Luis Potos\'i, Mexico}
\affiliation{$^{2}$ CONACYT - Unidad de Ciencias del Agua, Centro de 
Investigaci\'on Cient\'ifica de Yucat\'an A.C. (CICY), Calle 8, 
No. 39, Mz. 29, S.M.  64, 77524, Canc\'un, Quintana Roo, Mexico}
\affiliation{$^{3}$
Condensed Matter Physics Laboratory, Heinrich Heine Universit\"at,
Universit\"atsstra\ss{}e 1, 40225, D\"usseldorf, Germany}
\affiliation{$^{4}$ Departamento de F\'isica Aplicada, Cinvestav, Unidad 
M\'erida, Apartado Postal 73 Cordemex, 97310, M\'erida, Yucat\'an, 
Mexico}

\affiliation{$^{5}$ Departamento de Ingenier\'ia F\'isica, Divisi\'on
de Ciencias e Ingenier\'ias, Universidad de Guanajuato, Loma del Bosque
103, 37150, Le\'on, Mexico}
\affiliation{$^{6}$Department of Physics, Heinrich Heine Universit\"at
D\"usseldorf, Universit\"atsstra\ss{}e 1, 40225, D\"usseldorf, Germany}

\affiliation{$^{7}$ Institut f\"ur Materialphysik im Weltraum, Deutsches
Zentrum f\"ur Luft-und Raumfahrt (DLR), 51170, K\"oln, Germany}

\email{a)ramoncp@fisica.ugto.mx}
\email{b)luisfer.elizondo@gmail.com}
\date{\today}

\begin{abstract}

The binary hard-sphere mixture is one of the simplest representations of 
a many-body system with competing time and length scales. This model is 
relevant to fundamentally understand both the structural and 
dynamical properties of materials, such as metallic melts, colloids, 
polymers and bio-based composites. It also allows us to study how 
different scales influence the physical behavior of a multicomponent 
glass-forming liquid; a question that still awaits a unified description. 
In this contribution, we report on distinct dynamical arrest transitions 
in highly asymmetric binary colloidal mixtures, namely, a single 
glass of big particles, in which the small species remains ergodic, and 
a double glass with the simultaneous arrest of both components. 
When the mixture approaches any glass transition, the relaxation of the 
collective dynamics of both species becomes coupled. In the single glass 
domain, spatial modulations occur due to the structure of the large 
spheres, a feature not observed in the two-glass domain. The relaxation 
of the \emph{self} dynamics of small and large particles, in contrast, 
become decoupled at the boundaries of both transitions; the large 
species always displays dynamical arrest, whereas the small ones 
appear arrested only in the double glass. 
Thus, in order to obtain a complete picture of the distinct 
glassy states, one needs to take into account the dynamics of both 
species.

\end{abstract}
\maketitle

\section{Introduction}

The solidification of a liquid into an amorphous non-equilibrium glass 
is one of the most ubiquitous processes in nature, having also 
a remarkable scientific and technological relevance.
Metallic alloys \cite{greer}, bioactive materials \cite{zahid},
ceramics \cite{takahashi}, polymers \cite{hodge} and colloidal
suspensions \cite{cipelletti1,weitz,lu,zacarelli1,norman} are common 
examples of materials that under certain thermodynamic conditions are 
able to display a wide variety of non-crystalline structures 
and dynamically arrested states with solid-like properties that, 
so far, are not completely understood.

Acquiring full control of the physical properties of amorphous
materials, however, requires to fundamentally understand how the 
solid-like behavior appearing in an undercooled glass-forming
liquid is related to the dramatic slowing down of the microscopic 
dynamics \cite{angell,angell1}, and how such behavior is
mediated by both the direct interactions among the particles and the
thermodynamic conditions under which the material is prepared
\cite{weeks1,weeks2,weeks3}.

Model colloidal suspensions have played an important role in the study
of glasses and gels, since they provide access to time and length scales
at which most of the essential features of \emph{glassy} behavior are
manifested  \cite{vanmegen1,vanmegen2,sciortino,hunter}.
Colloids have noticeably improved our insight of the fundamental 
mechanisms for physical gelation \cite{lu,zacarelli1,norman} and the
glass transition \cite{cipelletti1,weitz}, providing neat
experimental realizations of dynamically arrested states in finely
controlled systems and conditions
\cite{vanmegen1,vanmegen2,jan,vanmegen,pham,pham1,eckert,marco1,marco2}.
One of the best known classes of colloidal model systems is the so-called 
\emph{hard-spheres} (HS) dispersion.
Despite their conceptual simplicity, colloidal HS are expected to retain
many important features of glass-forming liquids near the dynamical 
arrest transitions \cite{vanmegen1,vanmegen2,ameyer,thvg1}.
For a \emph{monodisperse} HS suspension, the formation of glassy states 
at volume fractions $\phi\gtrsim0.58$ is essentially due to a mechanism 
defined as ``\emph{caging}'', where the motions of individual particles 
are constrained by their nearest neighbors 
\cite{vanmegen1,vanmegen2, weeks}. The addition of a second species with 
a different size, however, modifies drastically this scenario:
for a narrow size disparity
($\delta\equiv\sigma_s/\sigma_b\gtrsim0.4$; $\sigma_s$ and $
\sigma_b$ being the diameters of the \emph{small} and \emph{big} 
particles, respectively), this implies a shift of the glass transition 
(GT) point to a larger total volume fraction 
\cite{vanmegen}, accompanied with spatial and temporal heterogeneities, 
and aging effects \cite{weeks1,weeks2,weeks3}.
For higher degrees of asymmetry, and depending on the composition, 
different and more complex glassy states, such as attractive and  
asymmetric glasses emerge \cite{jan,marco1,marco2, marco3,marco4,moreno,mayer,horbach}.

Attractive glasses and gels are observed in mixtures of HS
colloids and non-adsorbing polymers \cite{pham,pham1,eckert} and, 
so far, they have been described almost invariably in terms of an 
effective one-component system; the presence of the polymers results
in attractive depletion interactions among the colloidal particles
\cite{zacarelli1,lu,weitz,fabian,fuchs,dawson,nagi1}.
As we show in what follows, however, the glassy behavior of a
highly asymmetric binary mixture of HS can only be fully understood by
taking into account all the degrees of freedom of the two components, 
\emph{i.e.,} one should consider explicitly the dynamical contribution 
of both species. This suggests that a description based, for example, 
on the depletion forces acting on the large species can only provide a 
coherent picture for the static correlations of an \emph{effective} 
system composed of big attractive particles \cite{dijkstra,Erik2013}, 
but cannot offer any insight on the fundamental relevance that the 
dynamics of the smaller species possesses in the GT of the mixture.

From a theoretical point of view, the GT in asymmetric 
HS mixtures has been described by the mode coupling theory (MCT) 
\cite{goetze1,goetze2,goetze3,goetze4} and the self-consistent 
generalized Langevin equation (SCGLE) theory 
\cite{yeomans0,yeomans1,chavez,todos}. The predicted glassy
scenario \cite{rigo1,rigoAO,thvggotz,thomas} agrees with experimental 
observations for the large spheres \cite{marco1,marco2}.
Both approaches have predicted a richer and complex scenario for 
dynamical arrest, where the mobility of both species play a key role. 
Some of these features are in qualitative agreement with previous 
simulated results for binary mixtures of soft spheres 
\cite{moreno,mayer,horbach}.
However, to date no systematic and direct comparison between theory,
simulation and experiments has been presented to corroborate the
existence of different glassy states in highly asymmetric HS binary  
mixtures and to characterize in detail the \emph{self} and collective 
dynamics of both species towards the distinct dynamical arrest 
transitions. This is the main goal of the present work.

Thus, in this contribution we unravel the dynamical mechanisms that 
lead to the different arrested states and provide a unified 
description of glassy dynamics in this model system.
More specifically, we show how the glassy states differ 
in terms of the \emph{self} and collective dynamics of
both species, as well as the corresponding ergodicity parameters. 
We particularly emphasize the role of the small spheres, whose 
distinct relaxation patterns allow us to distinguish mixed glassy 
states (where only the large spheres undergo a GT), asymmetric glasses 
(where both species form a glass) and localization transitions of small 
particles in a glass of large spheres. We have
performed extensive \emph{event-driven} molecular dynamics (MD) 
simulations, developed
theoretical calculations within the SCGLE formalism and performed 
confocal differential dynamic microscopy (DDM) experiments 
A brief summary of the SCGLE theory, details of the simulations and 
experiments is provided in the supplemental material (SM).

\section{Arrested States Diagram of a Highly Asymmetric Binary Mixture
 of Hard-Spheres}\label{ddiagrams}

\begin{figure}
{\includegraphics[width=3.2in, height=3.2in]{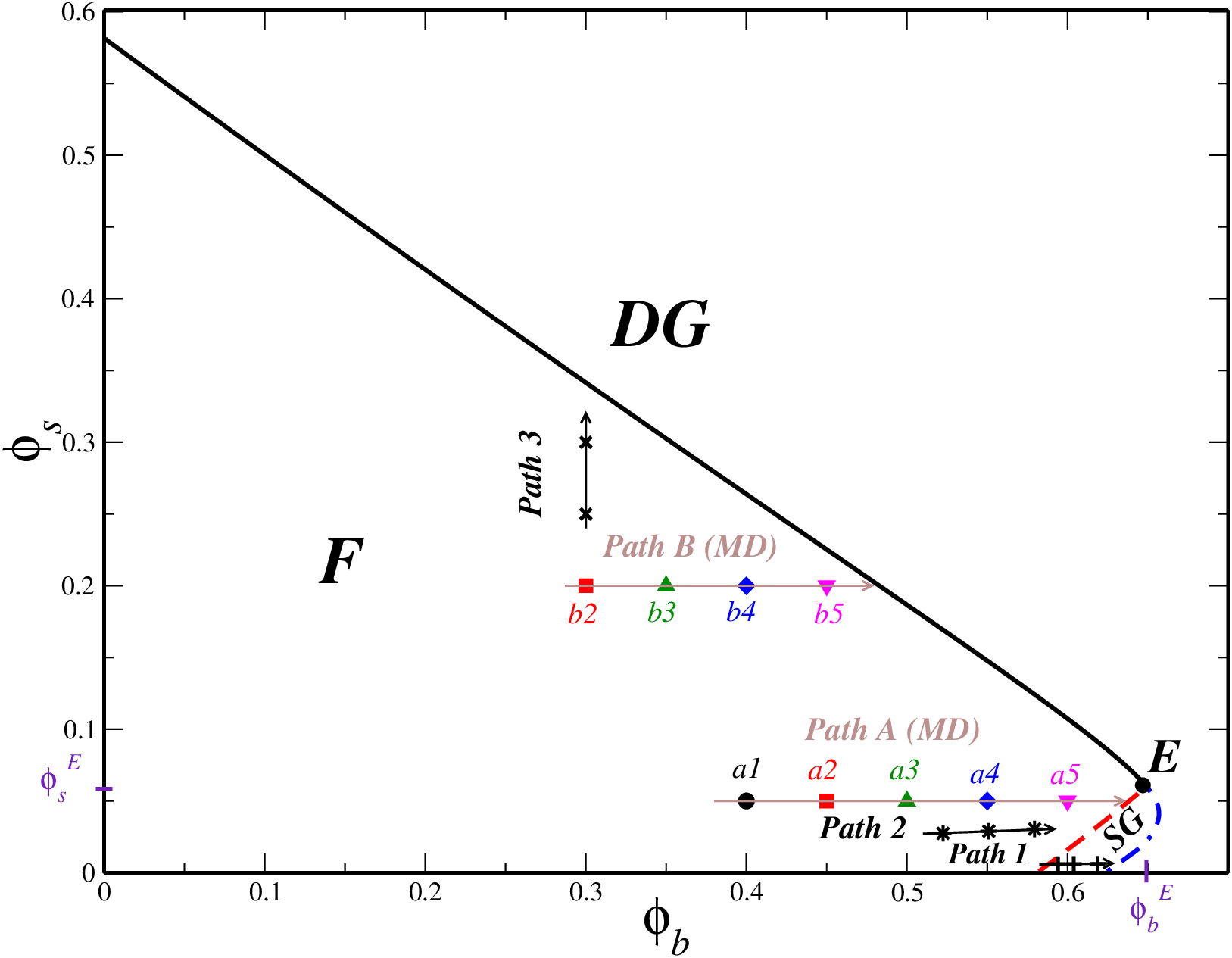}}
\caption{(Color online) Dynamical arrest diagram for a binary
mixture of HS with size ratio $\delta\approx0.2$,
as a function of the volume fractions of big, $\phi_b$, and 
small, $\phi_s$, spheres. Lines are predictions of the SCGLE 
theory \cite{rigo1} using the Percus-Yevick approximation \cite{percus}
combined with the Verlet-Weis correction \cite{verlet} for the
partial static structure factors. The (black) solid line indicates the
transition from the fluid (\textbf{\emph{F}}) to the \emph{double} glass
(\textbf{\emph{DG}}). The (red) dashed line indicates the transition 
between the \textbf{\emph{F}} and the \emph{single} glass
(\textbf{\emph{SG}}). The (blue) dashed-dotted line shows the
transition from the \textbf{\emph{SG}} to the \textbf{\emph{DG}}.
The three transition lines intersect at the state point 
$\bm{E}=(\phi_b^E,\phi_s^E)$.
Two different paths obtained at constant $\phi_s$ are shown:
\emph{Paths A} ($\phi_s=0.05$) and \emph{B} ($\phi_s=0.2$). The state
points $a1$-$a5$ and $b2$-$b5$ along each path represents compositions 
studied with MD simulations. \emph{Path} $1$ ($\boldsymbol{+}$)
corresponds to the experimental data of Ref. \cite{marco3}, whereas
\emph{Paths} $2$ ($\boldsymbol{\ast}$) and $3$ ($\boldsymbol{\times}$)
correspond to experimental data obtained in this work.} \label{Fig1}
\end{figure}

We first present the dynamical arrest diagram of largely asymmetric 
binary mixtures of hard spheres with a size ratio $\delta\approx 0.2$ 
as a function of the control parameters $(\phi_{b},\phi_{s})$, 
\emph{i.e.,} the volume fractions of, respectively, big and small 
spheres, as obtained by SCGLE (Fig. \ref{Fig1}) \cite{rigo1}. Here  
$\phi_{i}\equiv \frac{\pi} {6}\rho_{i}\sigma^{3}_{i}$ ($i=s,b$), 
where  $\rho_{i}\equiv N_{i}/V$ denotes the corresponding particle 
number density, and $N_{i}$ the number of particles of species 
$i$ in the volume $V$. For comparison, the diagram for 
$\delta\approx0.1$ and experimental data previously reported in Ref. 
\cite{marco2} are briefly presented and discussed in the SM.

Three different states can be distinguished: a fluid
(\textbf{\emph{F}}), where both species diffuse, a double glass
(\textbf{\emph{DG}}), in which both components are dynamically arrested, 
and a single glass (\textbf{\emph{SG}}) where the big particles 
undergo a GT, whereas the small particles diffuse through the voids left 
by large spheres, similar to the case of tracer particles in crowded 
\cite{marco3} and confining media \cite{porous}.
It has been shown previously that the region of mixed states,
\textbf{\emph{SG}}, expands when $\delta$ decreases, whilst it 
disappears for $\delta\gtrsim 0.38$ \cite{rigo1}. 
MCT provides qualitatively similar predictions \cite{thomas}.

The ergodic region \textbf{\emph{F}} is enclosed by two different 
transitions: The transition from a fluid to a single glass 
(\textbf{\emph{F}}-\textbf{\emph{SG}}, red dashed) and the
transition from a fluid to a double glass (\textbf{\emph{F}}-
\textbf{\emph{DG}}, black solid). 
The \textbf{\emph{F}}-\textbf{\emph{SG}} line runs from 
($\phi_b=0.582,\phi_s=0$), \emph{i.e.,} the GT point in the absence of 
small particles, to the \emph{crossing} point labeled as
\textbf{\emph{E}}$\equiv(\phi_b^E, \phi_s^E)$.
Along this line, $\phi_b$ monotonically increases with $\phi_s$.
This indicates that below $\phi_s^E$, a glass can be \emph{melted}
upon the addition of small particles  \cite{vanmegen}.
Above $\phi_s^E$, in contrast, one encounters the 
\textbf{\emph{F}}-\textbf{\emph{DG}} line, extending from 
\textbf{\emph{E}} to ($\phi_b=0,\phi_s=0.582$). 
Along this line, $\phi_b$ monotonically decreases as a function of 
$\phi_s$.
In the limit of small $\phi_b$, one observes a special type of
\emph{asymmetric} glass, where the large particles are localized
in a glass of small spheres, as already observed in previous work
\cite{marco1}.
Moreover, a third transition separates the single glass and double 
glass regions (\textbf{\emph{SG}}-\textbf{\emph{DG}}, blue 
dashed-dotted line). This transition describes the dynamical arrest 
of the small particles in the arrested large spheres.

For completeness, Fig. \ref{Fig1} also indicates the state points studied 
by means of MD simulations (\emph{Paths A} and \emph{B}) and DDM 
experiments (\emph{Paths} $1 (\boldsymbol{+}),2(\boldsymbol{\ast})$ and 
\textbf{$3(\boldsymbol{\times})$}), which are summarized in Table I.

\begin{center}
\begin{tabular}{l*{6}{c}r}
\emph{Path / Sample}  & $\phi_b$ & $\phi_s$ & SCGLE & MD& EXP & 
Transition \\
\hline
\it{A}	& 0.40-0.60 & 0.05 & \checkmark & \checkmark & $\times$ & $F-SG$ \\
$1.1$  & 0.594 & 0.006    & \checkmark & $\times$ & \checkmark & $F-SG$  \\
$1.2$  & 0.6039 & 0.0061  & \checkmark & $\times$ & \checkmark & $F-SG$  \\
$1.3$  & 0.6138 & 0.0062  & \checkmark & $\times$ & \checkmark & $F-SG$  \\
$2.1$  & 0.5225 & 0.0275 & \checkmark & $\times$ &  \checkmark & $F-SG$\\
$2.2$  & 0.551 & 0.029 & \checkmark & $\times$ &  \checkmark   & $F-SG$\\\vspace{0.5cm}
$2.3$  & 0.5795 & 0.0305 & \checkmark & $\times$ &  \checkmark & $F-SG$\\
\it{B}  & 0.35-0.45 & 0.20 & \checkmark & \checkmark & $\times$ & $F-DG$ \\
$3.1$   & 0.3 & 0.25 & \checkmark & $\times$ &  \checkmark & $F-DG$\\
$3.2$   & 0.3 & 0.30 & \checkmark & $\times$ &  \checkmark & $F-DG$\\
\end{tabular}\label{tab}
\captionof{table}{List of the state points studied by means of SCGLE
calculations, MD simulations and experiments for the transitions from 
the fluid (\textbf{\emph{F}}) to the single glass 
(\textbf{\emph{SG}}) and double glass (\textbf{\emph{DG}}).}
\end{center}

\section{Fluid-Glass Transitions (Predicted scenario)}

Let us first discuss the qualitative differences between the
\textbf{\emph{F}}-\textbf{\emph{SG}} and 
\textbf{\emph{F}}-\textbf{\emph{DG}} transitions on the basis of 
characteristic quantities, such as the localization lengths and 
the non-ergodicity parameters, as predicted by the SCGLE formalism.

\subsection{Localization length (self-dynamics)}
\label{loc_lengths}

\begin{figure}
\center
{\includegraphics[width=3.2in, height=3.2in]{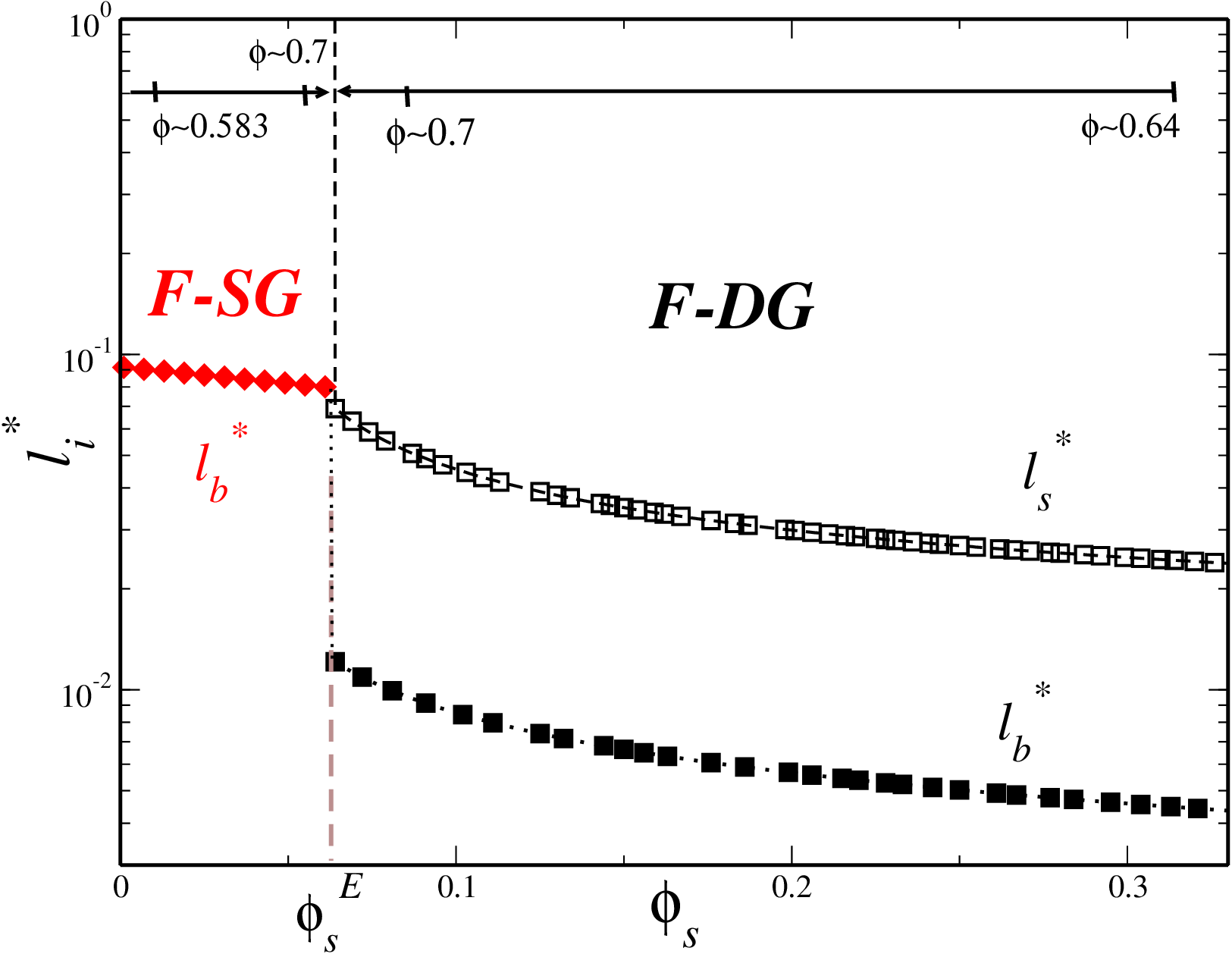}}
\caption{(Color online) Normalized localization lengths 
$l_b^*$ (solid symbols) and 
$l_s^*$ (open symbols) of the large and small 
particles, respectively, as function of the volume fraction of small 
particles, $\phi_s$, calculated with the SCGLE formalism along the 
two transition lines enclosing region \textbf{\emph{F}} in 
Fig. \ref{Fig1}.} \label{local}
\end{figure}

In an ordinary HS glass, the mean square displacement (MSD) shows a 
long-time plateau \cite{marco2}, whose height indicates the displacement 
inside a \emph{nearest-neighbors} cage. The square-root of this value is 
called localization length, $l_i$, and is a measure of the local 
confinement.
In Fig. \ref{local}, the behavior of the localization lengths
along the \textbf{\emph{F}}-\textbf{\emph{SG}} and 
\textbf{\emph{F}}-\textbf{\emph{DG}} lines is reported. For the 
\textbf{\emph{F}}-\textbf{\emph{SG}} transition 
($0\le\phi_s\le \phi_s^E$), the total volume fraction of the mixture,
$\phi=\phi_b+\phi_s$, increases when going from the point 
$(\phi_b=0.582,\phi_s=0)$ to \textbf{\emph{E}}, and the normalized 
localization length $l^*_{b}\equiv l_b/\sigma_b$ of the big spheres is 
found to be $l^*_b\approx 10^{-1}$. Hence, their characteristic 
\emph{cage size} corresponds to approximately 10\% of their diameter, a 
typical feature of an ideal glass of HS.
On the other hand, the normalized localization length of the small 
particles, $l^*_s\equiv l_{s}/\sigma_b$, is infinite,
indicating that the latter ones are not localized and 
diffuse (a feature not observed for a small degree of asymmetry 
\cite{weeks2}). These results suggest a dynamical decoupling of the 
\emph{self} dynamics of both species.

At the intersection of the \textbf{\emph{F}}-\textbf{\emph{SG}} 
and \textbf{\emph{F}}-\textbf{\emph{DG}} lines, \emph{i.e.,}
at \textbf{\emph{E}}, $l^*_b$ discontinuously jumps from 
$\sim10^{-1}$ to $\sim{10^{-2}}$ and $l^*_s$ becomes suddenly
finite with a value $l^*_s\sim{10^{-1}}$.
This indicates that along the \textbf{\emph{F}}-
\textbf{\emph{DG}} line, \emph{i.e.,} for $\phi_s>\phi_s^E$, both 
species are localized, with the small particles being less localized 
than the large ones. The latter become even more localized than along 
the \textbf{\emph{F}}-\textbf{\emph{SG}} transition 
\cite{marco2,marco1} (see also SM). 
In this case, the total volume fraction $\phi$ decreases when 
going from \textbf{\emph{E}} towards the point $(\phi_b=0,
\phi_s=0.582)$. Both $l^*_b$ and $l^*_s$ become smaller with 
decreasing $\phi$ and increasing $\phi_s$, which implies 
that the cage size along the \textbf{\emph{F}}-\textbf{\emph{DG}} line 
is controlled by the small spheres.

\subsection{Non ergodicity parameter (collective dynamics)}

\begin{figure}
\center
{\includegraphics[width=3.2in, height=2.1in]{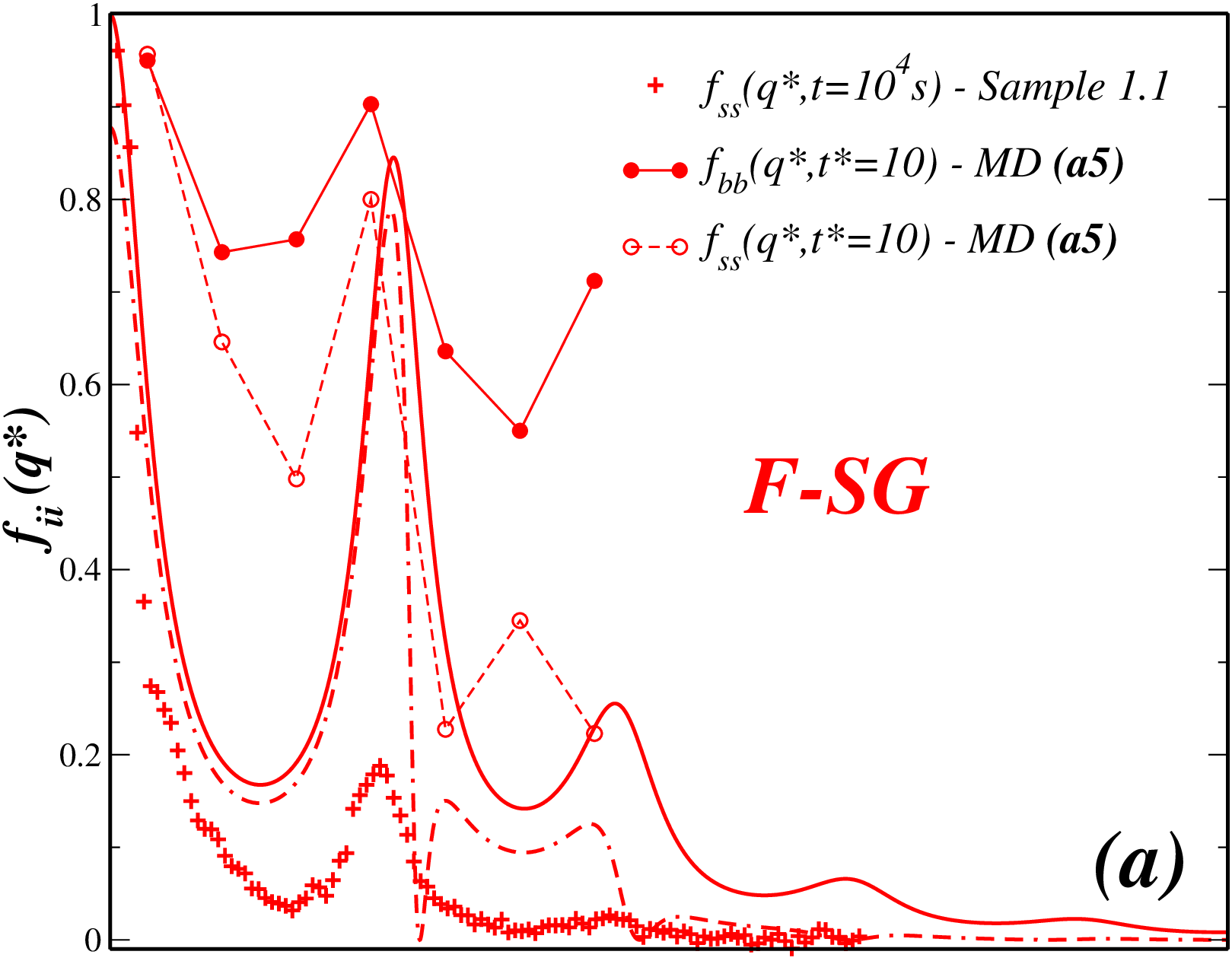}}\\
{\includegraphics[width=3.2in, height=2.1in]{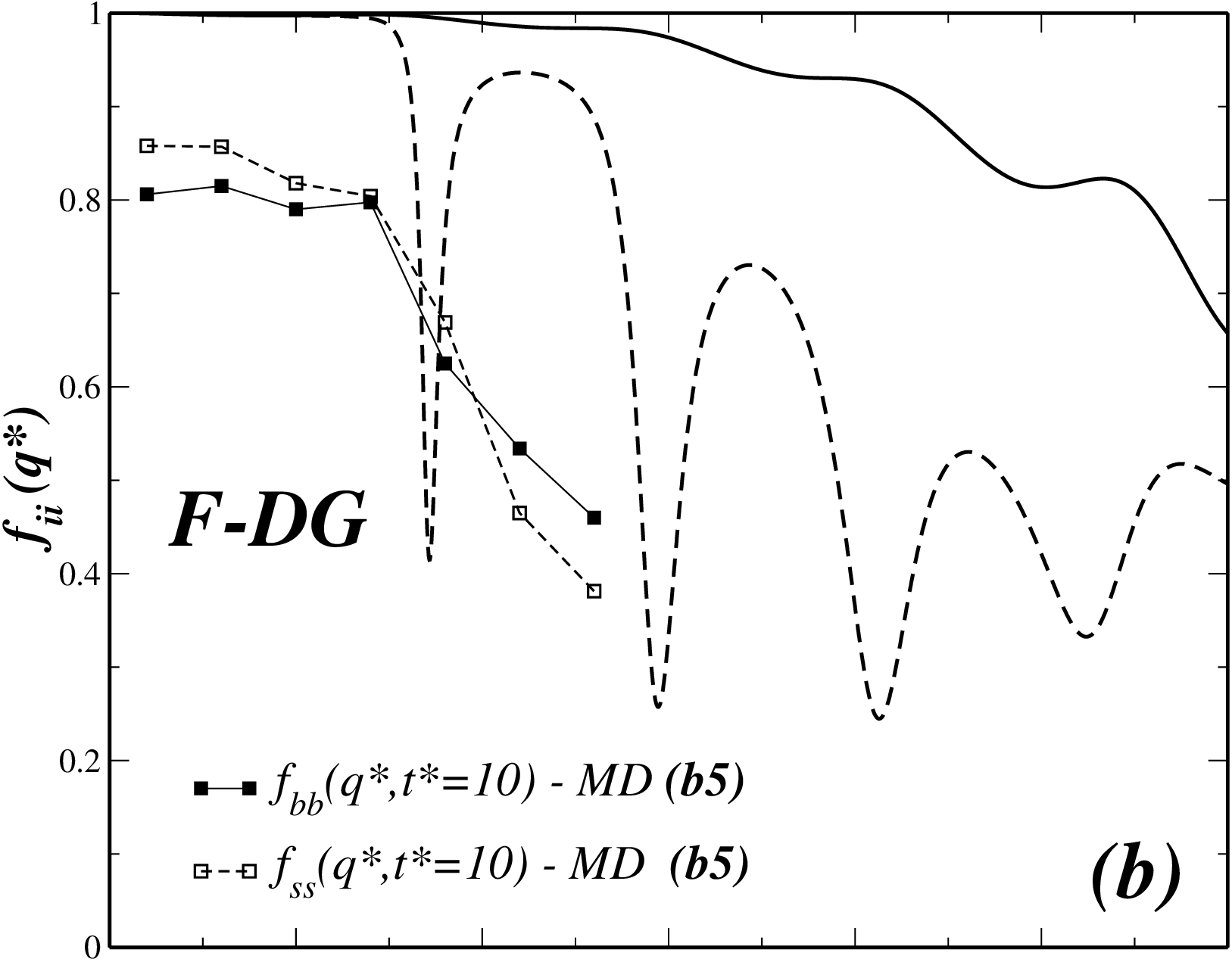}}\\
{\includegraphics[width=3.2in, height=2.1in]{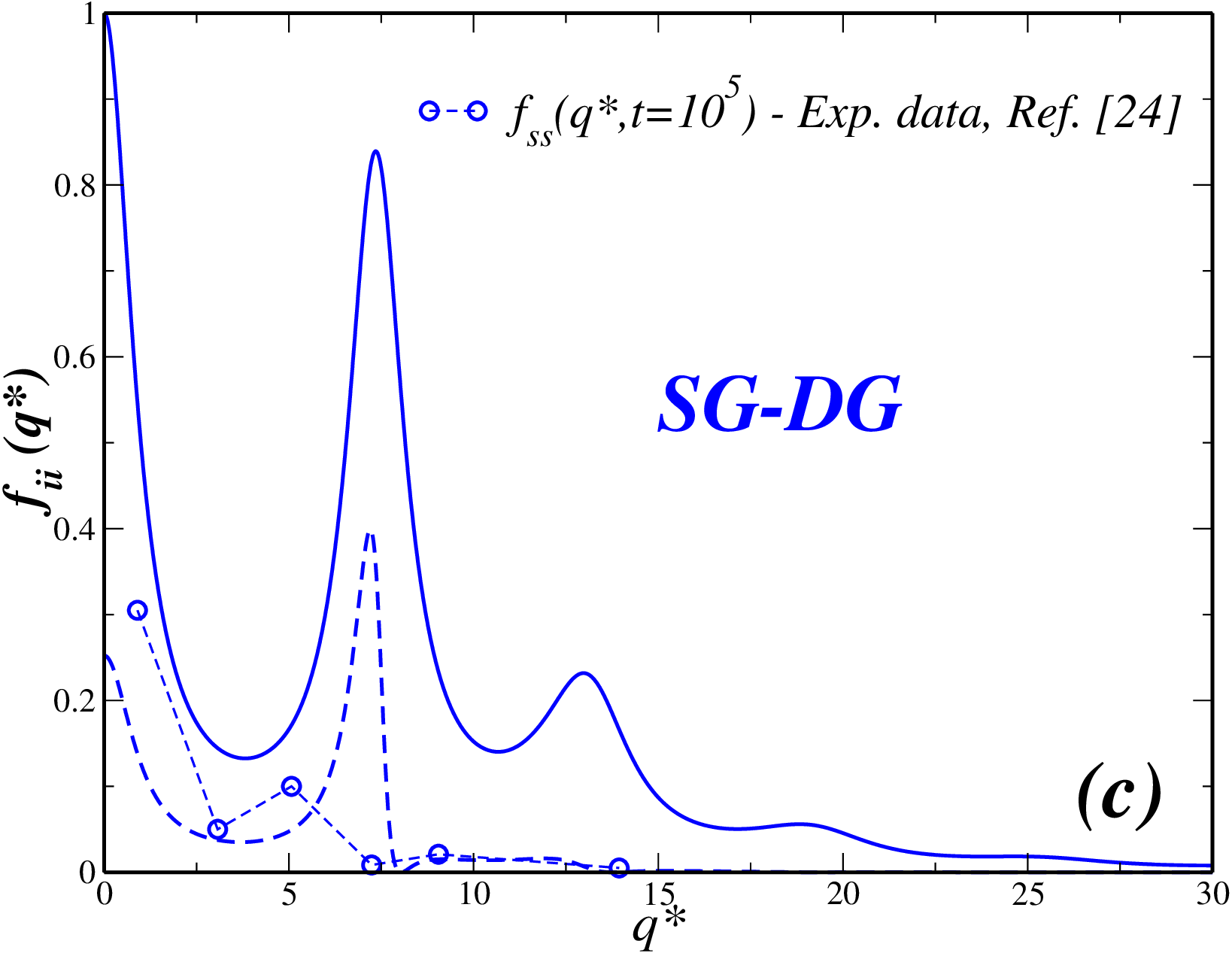}}
\caption{(Color online) \emph{Non-ergodicity} parameter (NEP) of the 
large particles, $f_{bb}(q^*)$ (solid lines), and of the small 
particles, $f_{ss}(q^*)$ (dashed-lines), as a function of the reduced 
wavenumber $q^{*}$ as predicted by the SCGLE formalism: 
\emph{(a)} at the intersection of \emph{Path A} with the
\textbf{\emph{F}}-\textbf{\emph{SG}} transition line; representative 
results extracted from the MD simulations at the state point $a5$
and experimental data from Ref. \cite{marco3} (Sample $1.1$)
are also displayed, \emph{(b)} at the intersection of \emph{Path B} 
with the \textbf{\emph{F}}-\textbf{\emph{DG}} transition line; 
representative results from the MD at the state point $b5$ are shown, 
\emph{(c)} at the intersection of \emph{Path 1} with the 
\textbf{\emph{SG}}-\textbf{\emph{DG}} transition line.} \label{ergo}
\end{figure}

Another fundamental difference between the two transitions enclosing
region \textbf{\emph{F}} is characterized by the so-called \emph{non 
ergodicity} parameters (NEP) \cite{vanmegen1,vanmegen2}
$f_{ii}(q)\equiv\displaystyle{\lim_{t\to\infty}F_{ii}(q,t)/S_{ii}(q)}$,
\emph{i.e.,} the \emph{nondecaying} components of the collective
intermediate scattering functions (ISFs)
$F_{ii}(q,t)=\langle\sum_{j,k}^{N}\exp(i\mathbf{q}\cdot[\mathbf{r}^{(i)}
_j(t)-\mathbf{r}^{(i)}_k(0)])\rangle/N$, where $\mathbf{q}$ is the 
scattering vector, $\mathbf{r}^{(i)}_j(t)$ describes the position of the 
$j$-th particle of species $i$ at time $t$, and $S_{ii}(q)$ denotes the 
corresponding partial static structure factor 
($S_{ii}(q)=F_{ii}(q,t=0)$).
The emergence of non-zero values for $f_{ii}(q)$ is associated with the 
dynamical arrest of the collective dynamics of 
species $i$.

Fig. \ref{ergo} shows the predictions of the SCGLE theory for the
NEP of the big , $f_{bb}(q^*)$, and small, $f_{ss}(q^*)$, particles at
three points located on the 
(\emph{a}) \textbf{\emph{F}}-\textbf{\emph{SG}},
(\emph{b}) \textbf{\emph{F}}-\textbf{\emph{DG}}  and
(\emph{c}) \textbf{\emph{SG}}-\textbf{\emph{DG}} lines as a
function of the reduced wave number $q^* \equiv q\sigma_b$. In order to 
later compare with the simulation results, we have chosen the points on 
the \textbf{\emph{F}}-\textbf{\emph{SG}}
and \textbf{\emph{F}}-\textbf{\emph{DG}} lines that correspond to the
crossing points with the two paths investigated with MD, indicated
as \emph{Paths A} and \emph{B}. Similarly, we have
considered the \emph{Sample 1.1} (see Table I), which almost intersects 
the transition line \textbf{\emph{F}}-\textbf{\emph{SG}}. 

For the point at the intersection  of \emph{Path A} with the
\textbf{\emph{F}}-\textbf{\emph{SG}} transition line  (Fig. \ref{ergo}
(\emph{a})), one observes an oscillatory behavior in both 
$f_{bb}(q^*)$ and $f_{ss}(q^*)$, associated with modulations of the 
structure factor of the big species, $S_{bb}(q^*)$. Both 
NEPs appear coupled and they  are essentially identical up to 
$q^*\approx7.18$, which approximately 
corresponds to the location of the main peak of $S_{bb}(q^*)$. 
Thus, at large length scales, the collective dynamics of the small 
spheres is controlled by the confinement of the large particles.
For $q^*>7.18$, oscillations are still present, but become  
decoupled. The NEPs cease to oscillate and decay to nearly zero
at values $q^*\approx20$ in the case of $f_{ss}(q^*)$ and $q^*\approx30$
for $f_{bb}(q^*)$. 
This indicates that, at smaller length scales, the small spheres can 
explore the local environment. 
Data for both NEPs obtained from MD simulations at long time ($t^*=10$), 
and for $f_{ss}(q^*)$ obtained from DDM experiments (at $t=10^4s$) shows 
also oscillations at comparable $q^{*}$ values. This will be discussed in 
detail below (see, for instance, Figs. \ref{qdependence} and
\ref{path1}).  

In contrast, for the point at the intersection of \emph{Path B} with the
\textbf{\emph{F}}-\textbf{\emph{DG}} transition line  
(Fig. \ref{ergo}(\emph{b})), one observes a different behavior.
Both $f_{bb}(q^*)$ and $f_{ss}(q^*)$ appear coupled for $q^*\leq7.18$, 
but now remain constant at about unity. For $q^*>7.18$, the NEPs are decoupled and large oscillations appear only in $f_{ss}(q^*)
$. Also, larger \emph{spectra} of non-decaying components 
in the ISFs of both species are observed up to $q^*\approx150$ (not 
shown).  NEPs obtained from MD simulations are also constant for 
$q^*<7.18$, but decay for larger $q^*$ and appear coupled. This will also
be discussed below (see Fig. \ref{qdependence2}).

In Fig. \ref{ergo}(\emph{c}), the SCGLE predictions 
for the behavior of the NEPs at the intersection of the 
\textbf{\emph{SG}}-\textbf{\emph{DG}} transition line with \emph{Path 1}
are shown. The behavior is qualitatively similar to that observed at the
\textbf{\emph{F}}-\textbf{\emph{SG}} transition, but $f_{ss}(q^*)$ is
noticeable smaller than $f_{bb}(q^*)$ for $q^*\leq7.18$, and becomes
nearly  zero for $q^*>7.18$ (note that along \emph{Path A}
$\phi_s=0.05$, whereas along \emph{Path 1} $\phi_s\approx0.0062$). 
Hence, at this transition the small particles become trapped in voids 
created by big particles, thus resembling a localization transition in 
random porous media \cite{porous}.

\section{Dynamics from the Fluid to the Single Glass}\label{FSG}

We have investigated the dynamics of the mixture towards the 
\textbf{\emph{F}}-\textbf{\emph{SG}} transition
by means of theory, simulations and experiments. We have carried out 
\emph{event-driven} MD simulations with $\delta=0.2$ and following 
\emph{Path A}, \emph{i.e.}, fixing $\phi_s=0.05$ and varying $\phi_b$. 
In the experiments, we have studied mixtures with $\delta\approx0.18$
and followed two different tr
ajectories at fixed composition of the 
small species, namely, $x_s\equiv\phi_s/
\phi=0.01$ (\emph{Path} $1$) and $x_s=0.05$ (\emph{Path} $2$), and 
varied the total volume fraction $\phi$. Details of the simulations and 
experiments are provided in the SM and Ref. \cite{edilio}.

\subsection{Self dynamics at the \textbf{\emph{F}}-\textbf{\emph{SG}} transition}
\begin{figure}
{\includegraphics[width=3.2in, height=3.2in]{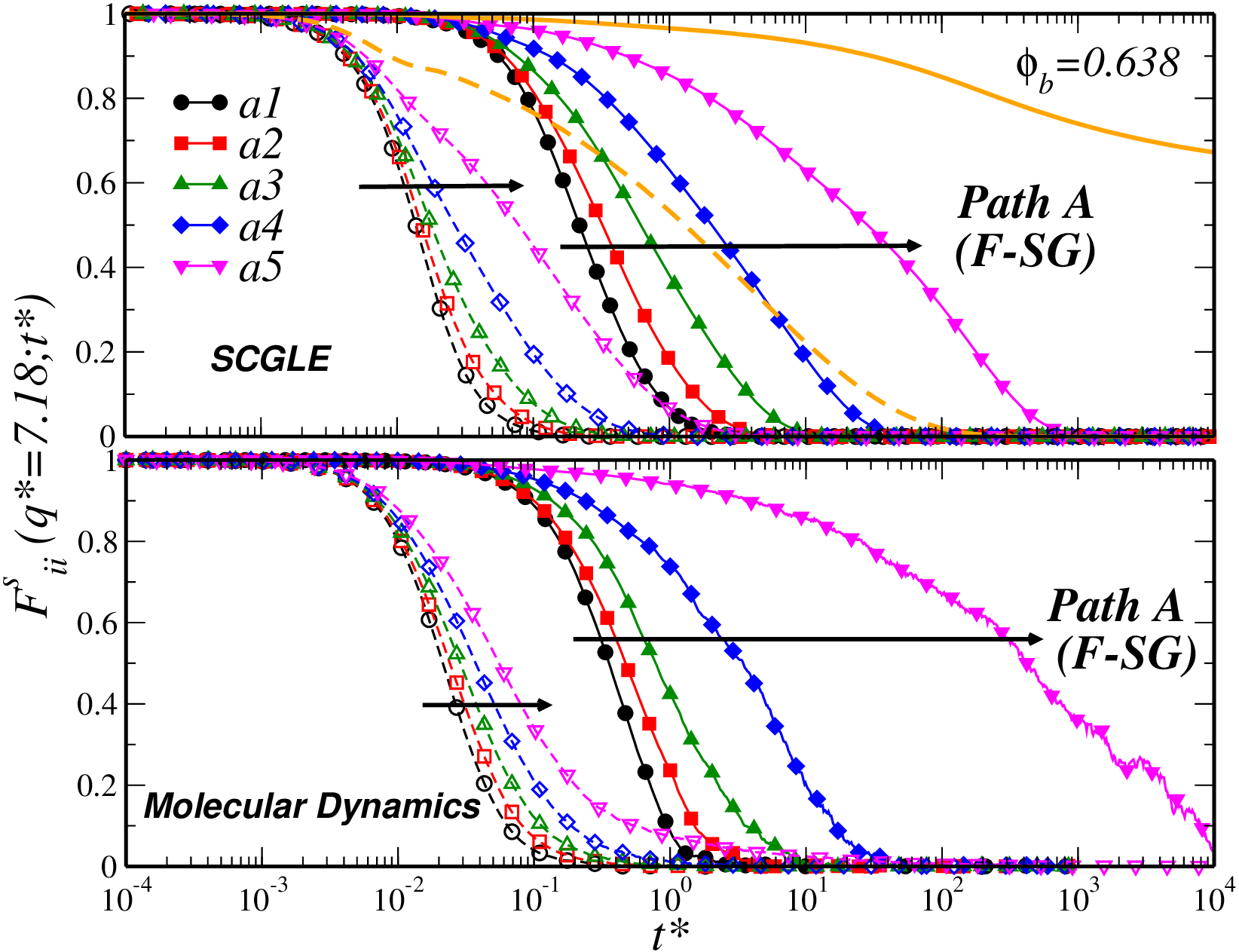}}
\caption{(Color online) Self ISFs for the large, 
$F^{S}_{bb}(q^{*},t^{*})$ (solid symbols), and small particles, 
$F^{S}_{ss}(q^{*},t^{*})$ (open symbols), 
calculated along \emph{Path A} (as indicated) at fixed $q^*=7.18$ and 
as a function of the reduced time $t^*\equiv t/t^0_b$, where 
$t^0_b=\sigma_b\sqrt{M_b/k_BT}$, $M_b$ is the mass of any of the large 
particles, $k_{B}$ the Boltzmann constant and $T$ the absolute 
temperature (Ref. \cite{edilio} and SM).
The upper panel displays the results of the SCGLE theory
(including the (orange) solid and dashed curves for, respectively, the 
big and small particles, at the critical value $\phi^{(g)}_b=0.638$)
and the lower panel shows results of MD simulations.} \label{fselfA}
\end{figure} 

Fig. \ref{fselfA} displays a comparison between the SCGLE
predictions (upper panel) and MD results (lower panel) for the behavior
of the \emph{self} ISFs,  $F^{S}_{ii}(q,t)\equiv
\langle\exp{[i\mathbf{q}\cdot\Delta\mathbf{R}^{(i)}(t)]}\rangle$, along
\emph{Path A} and evaluated at $q^{*}=7.18$;  where
$\Delta\mathbf{R}^{(i)}(t)$ denote the displacement of \emph{any} of 
the $N_{i}$ particles of species $i(=b,s)$ over a time $t$.
To allow for a one-to-one comparison between the SCGLE results and the 
MD simulations, we have appealed to the \emph{molecular} version of the 
SCGLE theory \cite{edilio}.

The \emph{self} ISFs obtained from SCGLE decay much faster for the 
small than for the large spheres. Upon increasing $\phi_b$, a general 
slowing down is observed with the difference between the relaxation 
times of small and big particles strongly increasing. 
Nevertheless, the \emph{self} ISF of the small species always decays 
to zero, as expected (Sec. \ref{loc_lengths}). In contrast, the
relaxation of $F^S_{bb}(q^*=7.18;t)$ becomes much slower and, at the
critical volume fraction $\phi_b^{(g)}=0.638$ (orange solid line), it 
develops the characteristic two-steps relaxation in the dynamics (this 
state point has no counterpart in the simulations). Thus, the only 
signature of dynamical arrest appears for the large species.

The above scenario qualitatively agrees with the MD simulations 
(lower panel of Fig. \ref{fselfA}), although with some differences.
For instance, at the state point $a5$, both theory and simulations show 
a pronounced decoupling. However, the MD results reveal a  
slower relaxation of the large spheres and a slightly different final 
relaxation of the small ones.
The latter reminds on the dynamical landscape of tracers in crowded 
environments, for instance, diffusion in heterogeneous porous media 
\cite{porous}. 
At this stage, we should point out that the MD simulations at $a5$ 
disclosed the structure of the large particles similar to that of a 
highly amorphous material, see, e.g., SM, which nevertheless exhibits 
a particle dynamics reminiscent of a GT.

\begin{figure}
{\includegraphics[width=3.2in, height=3.2in]{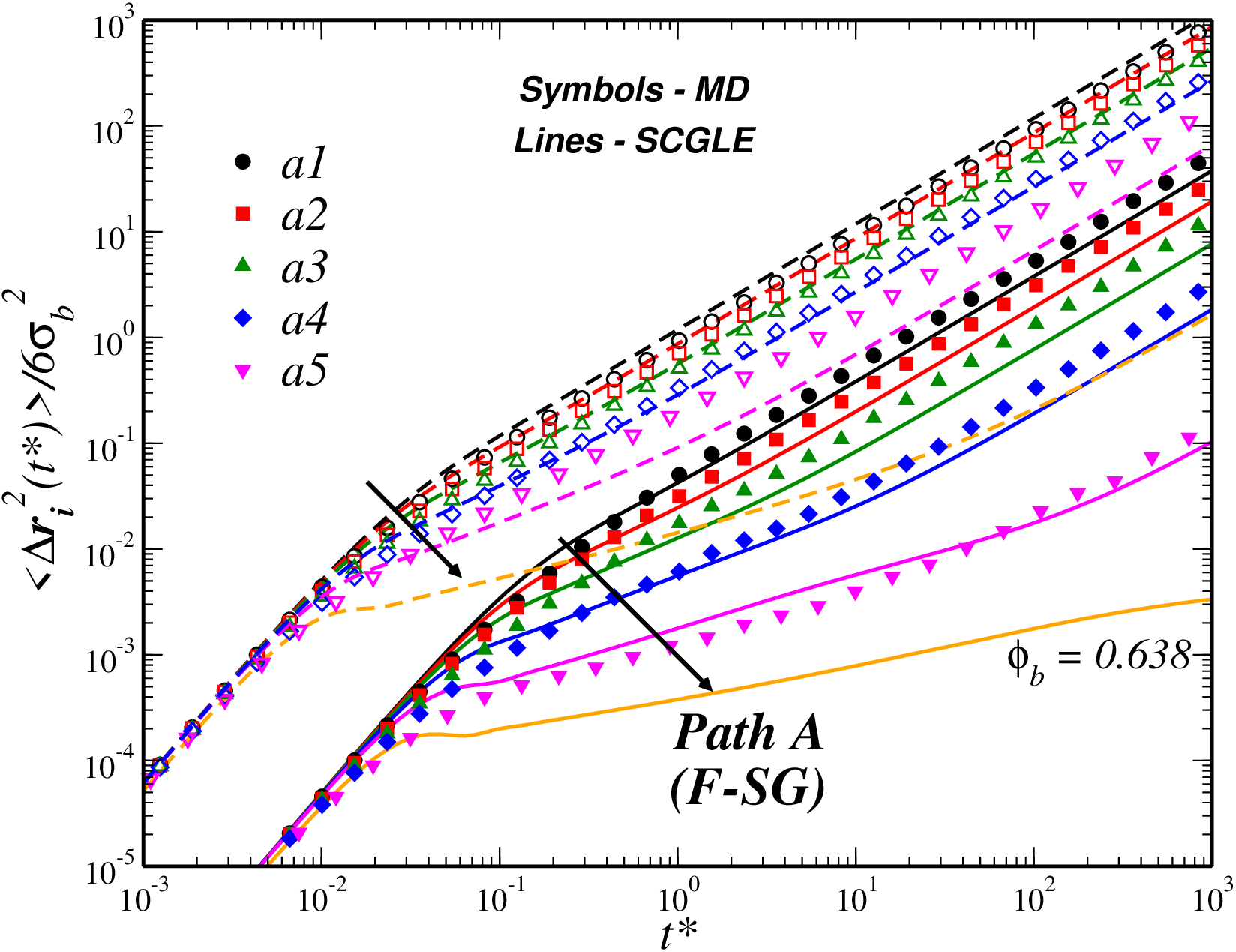}}
\caption{(Color online) MSDs for the large (solid symbols
and solid lines) and small particles (open symbols and dashed lines)
along \emph{Path A} (as indicated) obtained from MD simulations (symbols) 
and the SCGLE theory (lines).}
\label{msdsA}
\end{figure}

Fig. \ref{msdsA} presents the corresponding MSDs obtained from MD 
and SCGLE. The features of the \emph{self} ISFs are also manifested in 
the corresponding MSDs
$W_i^*(t;\phi_b,\phi_s)\equiv \langle(\Delta \textbf{\textit{r}}
_i(t))^2\rangle/6\sigma_b^2$, $(i=s,b)$.
For instance, at intermediate and long times, the MSD of the smaller 
particles decreases approximately one order of magnitude when 
increasing $\phi_b$ along \emph{Path A}, whereas the MSD of the large 
particles decreases by more than two orders of magnitude 
and exhibits an increasingly extended subdiffusive regime  at 
intermediate times. This is in agreement with the
physical scenario outlined in Fig. \ref{local}.

\subsection{Collective dynamics at the \textbf{\emph{F}}-\textbf{\emph{SG}} transition}
\begin{figure}
{\includegraphics[width=3.2in, height=3.2in]{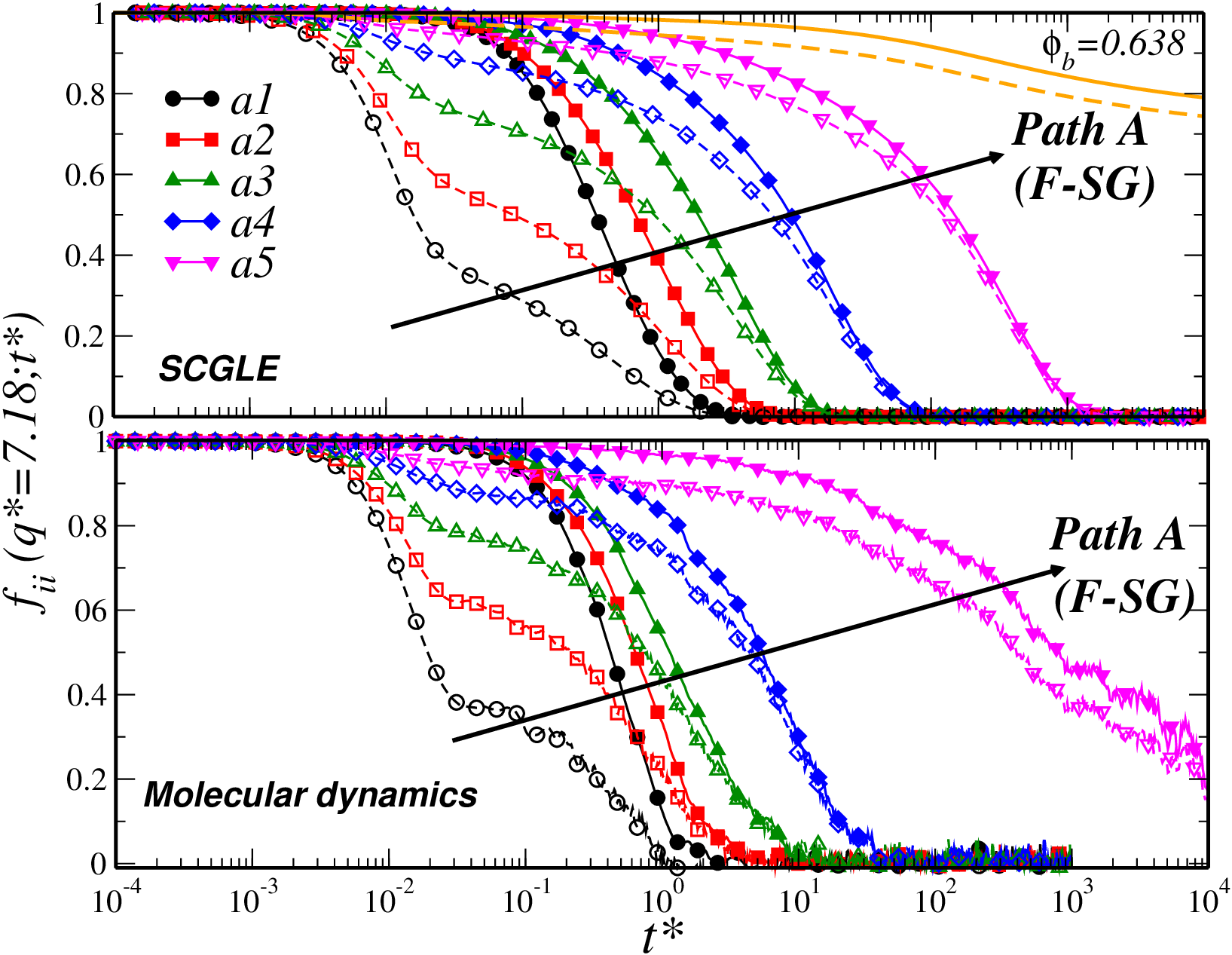}}
\caption{(Color online) Normalized collective ISFs for the large, 
$f_{bb}(q^{*},t^{*})$ (solid symbols), and small particles, $f_{ss}
(q^{*},t^{*})$ (open symbols), calculated along \emph{Path A} (as 
indicated) at fixed $q^*=7.18$ as a function of the reduced time $t^*$. 
The upper panel displays the results from the SCGLE theory and the lower 
panel from MD simulations.} \label{fcoleA}
\end{figure}

We now turn to the \emph{normalized} collective ISFs, 
$f_{ii}(q,t)\equiv F_{ii}(q,t)/S_{ii}(q)$. 
Fig. \ref{fcoleA} displays the predictions of the SCGLE (upper panel)
and MD data (lower panel) for $f_{ii}(q^*=7.18,t^*)$ along \emph{Path A},
which both provide essentially the same scenario.  
The behavior of $f_{bb}$ is very similar to $F^{S}_{bb}$ (see Fig. 
\ref{fselfA}). Instead, $f_{ss}$ noticeably differs from $F^S_{ss}$. 
At the state point $a1$, for instance, the former exhibits a two-step 
relaxation not observed in the latter. Upon increasing $\phi_b$, however, 
$f_{ss}$ gradually evolves and eventually follows the same trend as 
$f_{bb}$. Thus, approaching the \textbf{\emph{F}}-\textbf{\emph{SG}} 
transition the \emph{normalized} collective ISFs of both species become 
slower and coupled at wavenumbers $q^*=7.18$, as discussed above 
(Fig. \ref{ergo}(a)). These features are qualitatively the same in both 
SCGLE and MD results, although the latter show a slower relaxation with 
respect to the former at the state point $a5$.

A few remarks might be in order. As mentioned before, the 
\textbf{\emph{SG}} domain describes a region of 
partially arrested states, where the large spheres are predicted to 
undergo dynamical arrest and the small ones remain ergodic. 
At the level of \emph{self} diffusion quantities, this feature of the 
\textbf{\emph{F}}-\textbf{\emph{SG}} transition was
observed through the decoupling of the self ISFs, $F_{bb}^S$ and 
$F_{ss}^S$, and of the MSDs, $W_{b}(t)$ and $W_{s}(t)$ 
(Figs. \ref{fselfA} and \ref{msdsA}, respectively). 
At the level of collective dynamics, however, there is a subtle 
feature that deserves to be briefly commented.
To describe collective diffusion, one conventionally considers the 
\emph{normalized} ISFs, $f_{bb}$ and $f_{ss}$, which are quantities
accessible experimentally. According to Fig. \ref{fcoleA}, both ISFs 
display a slowing down in the relaxation and become strongly correlated 
towards the \textbf{\emph{F}}-\textbf{\emph{SG}} transition, thus 
suggesting the deceiving notion that, at the level of collective 
variables, it is only possible to detect either fluid  or arrested states 
but not partially arrested ones.
Nonetheless, this is only the result of the convention adopted to  
describe collective diffusion. 
To see this, one may consider the \emph{propagator} matrix 
$\Psi(q,t)\equiv F(q,t)S^{-1}(q)$, with initial condition 
$\Psi(q,t=0)=I$. In terms of the diagonal \emph{propagators} 
$\Psi_{bb}(q,t)$ and $\Psi_{ss}(q,t)$, the scenario for the collective 
dynamics of the \textbf{\emph{F}}-\textbf{\emph{SG}} transition differs 
from that displayed by $f_{bb}(q,t)$ and $f_{ss}(q,t)$.
Specifically, one finds that the collective propagator 
$\Psi_{bb}(q^*=7.18,t)$ displays dynamical arrest, whereas 
$\Psi_{ss}(q^*=7.18,t)$ decays to zero, in qualitative similitude with 
the behavior of the \emph{self} ISFs (for a detailed discussion, the 
reader is referred to Sec. IV of Ref.\cite{rigo1} and the SM).

\begin{figure}
{\includegraphics[width=3.2in, height=3.2in]{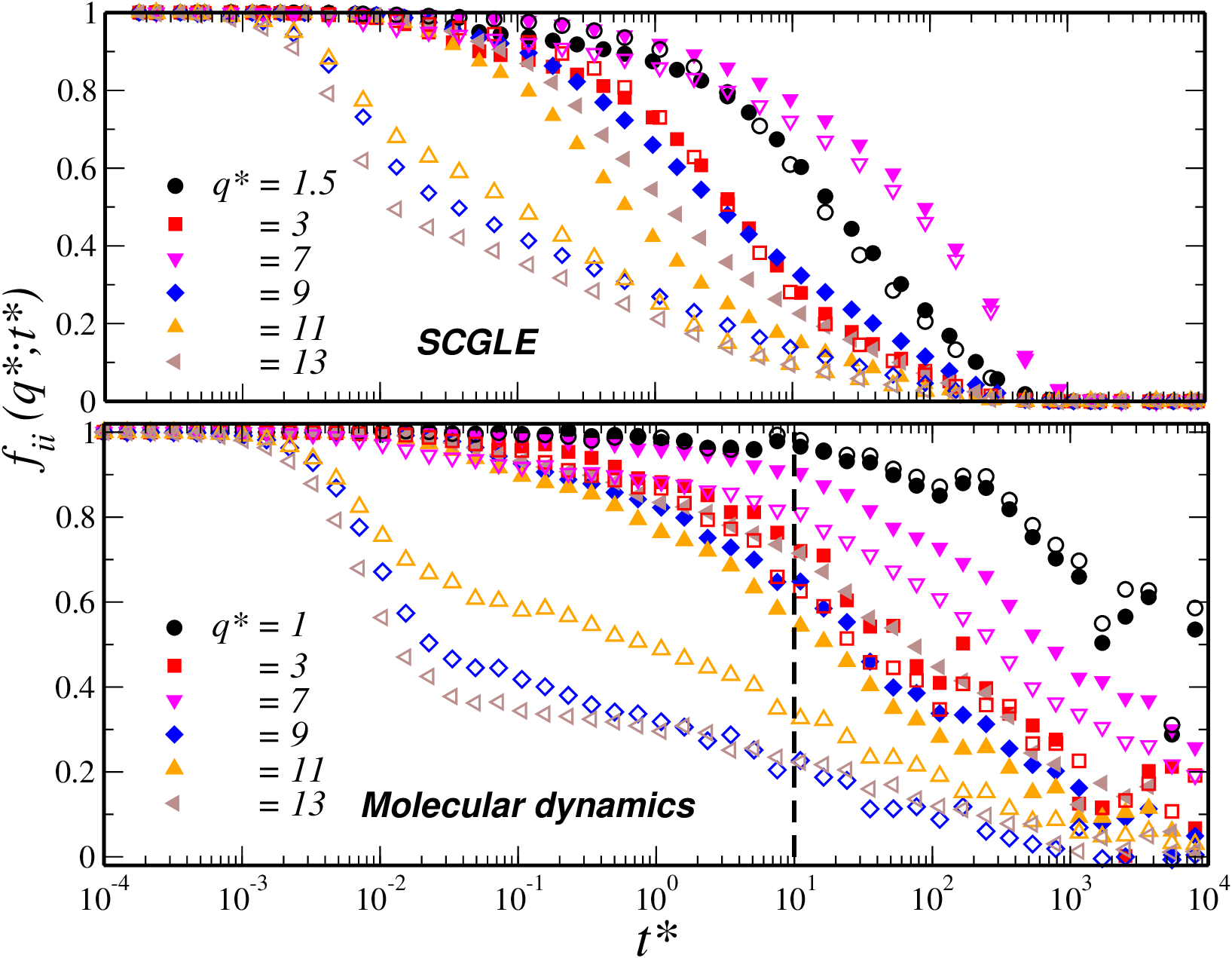}}
\caption{(Color online) Normalized collective ISFs for the large,
$f_{bb}(q^*,t^*)$ (solid symbols), and small particles, 
$f_{ss}(q^*,t^{*})$ (open symbols), at the state point
$a5=(\phi_b=0.60,\phi_s=0.05)$, for different $q^*$ (as indicated) 
as a function of the reduced time, $t^*$ obtained from the SCGLE 
theory (upper panel) and MD (lower panel). The vertical dashed line in 
the lower panel indicates $t^*=10$, at which the values 
$f_{ii}(q^*,t^*=10)$ have been extracted for the comparison in 
Fig. \ref{ergo}.}
\label{qdependence}
\end{figure}

We have additionally analyzed the $q^*$-dependence of $f_{ii}(q^*,t^*)$ 
for the sample closest to the \textbf{\emph{F}}-\textbf{\emph{SG}} 
transition along \emph{Path A}, \emph{i.e.,} the state point $a5$. 
As shown in Fig. \ref{qdependence}, both $f_{bb}$ and $f_{ss}$ display an 
initial acceleration of the decay with increasing $q^*$, a slowing down 
for $q^*= 7$, and a second acceleration for larger $q^*$. 
These effects are attributable to the modulation of the structure 
factor of the large particles, as already mentioned (Fig. \ref{ergo}(a)). 
Additionally, for $q^*>7$, $f_{ss}$ develops distinct relaxation 
patterns as those observed in $f_{bb}$; a faster initial decay followed 
by an intermediate inflection point, whose height \emph{oscillates} with 
increasing $q^*$.
This reflects the increasingly smaller fraction of small particles that 
are temporarily trapped at increasingly shorter length scales 
(larger $q^*$ values). 
\begin{figure}
{\includegraphics[width=3.2in, height=3.9in]{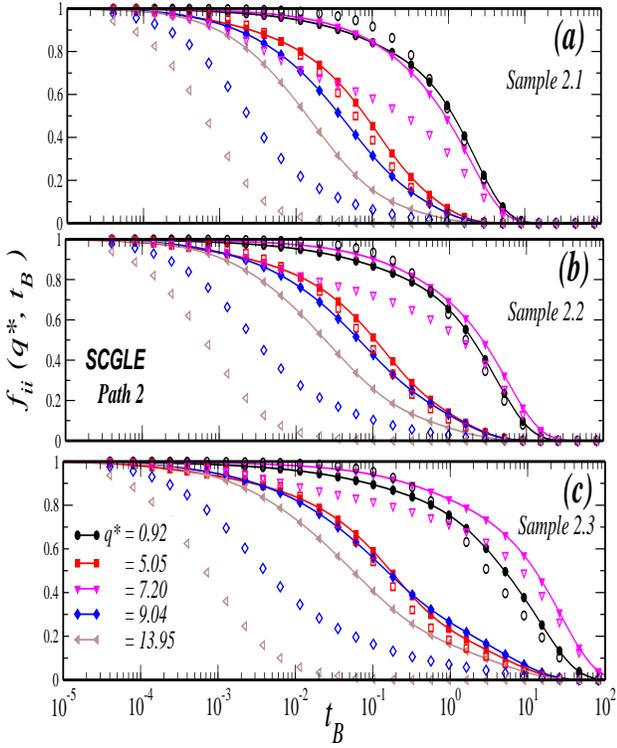}}
\caption{(Color online) Normalized collective ISFs for the large,
$f_{bb}(q^*,t_B)$ (solid symbols), and small particles, 
$f_{ss}(q^*,t_B)$ (open symbols), along \emph{Path 2} (Samples 
(a)$2.1$, (b)$2.2$, (c)$2.3$) for different wave numbers $q^*$ 
(as indicated) as a function of time $t_B\equiv t/D^0_b\sigma_b^{-2}$  
predicted by the SCGLE theory.}
\label{path2_theory}
\end{figure}

\begin{figure}
{\includegraphics[width=3.2in, height=3.8in]{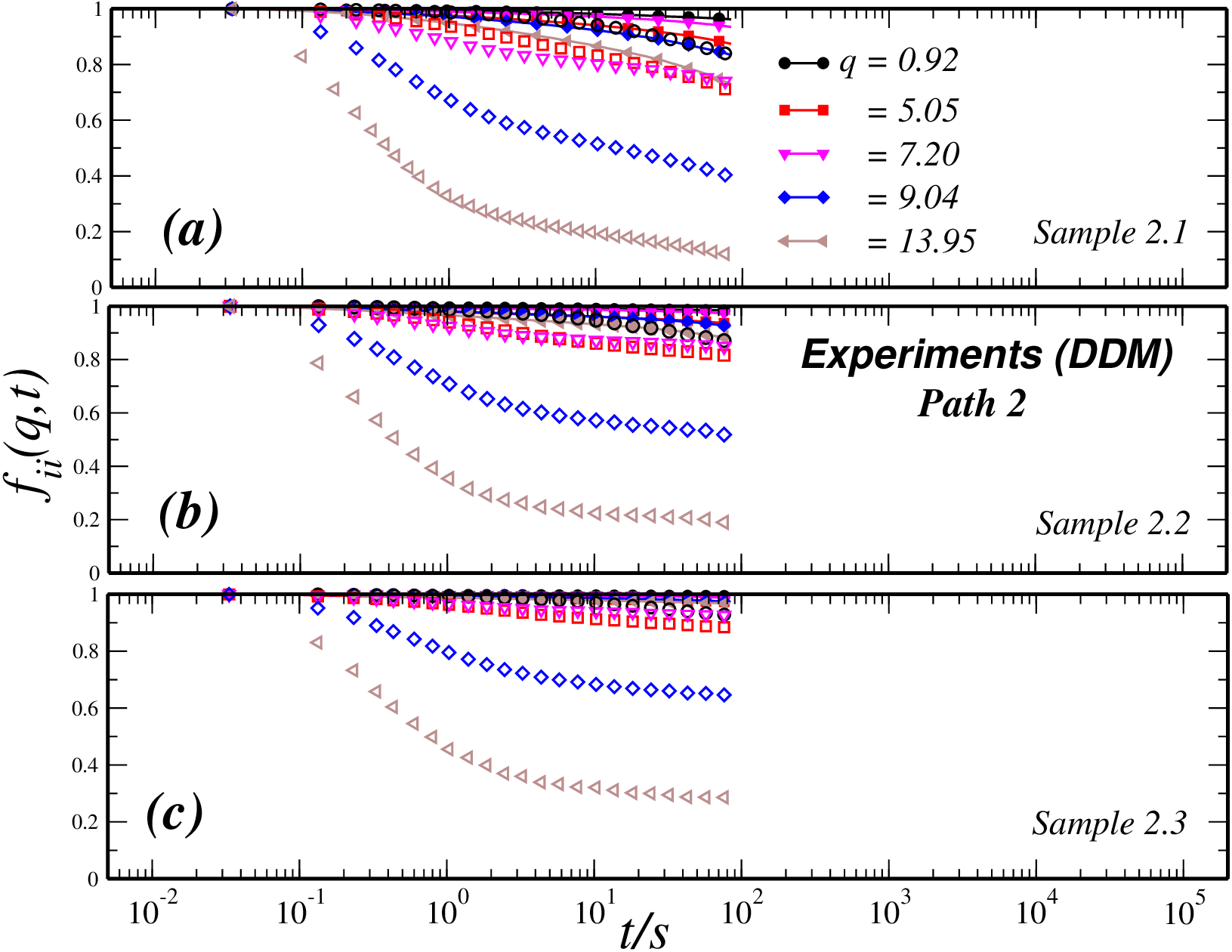}}
\caption{(Color online) Normalized collective ISFs for the large, 
$f_{bb}(q^*,t)$ (solid symbols), and small particles, $f_{ss}
(q^*,t)$ (open symbols), along \emph{Path 2} (Samples 
(a)$2.1$, (b)$2.2$, (c)$2.3$) for different wave numbers $q^*$
(as indicated) as a function of time $t$ as observed in experiments.}
\label{path2_exp}
\end{figure}

To test the influence of $\phi_s$ on the $q^{*}$-dependence of the 
dynamics, theoretical and experimental data along \emph{Path 2}  
($\phi_s \approx 0.03$) are reported in Figs. \ref{path2_theory} 
and \ref{path2_exp}, respectively.
The SCGLE predicts a similar scenario as that found for $\phi_s = 0.05$, 
involving a non-monotonic behavior in the relaxation of both species
upon varying $q^*$.
Despite the smaller time resolution of the experiments, the 
measured ISFs show similar trends. The SCGLE and experimental results 
suggest different decay patterns and transient plateaus of $f_{ss}$ 
for $q^*\approx9$, although with some quantitative differences that 
become larger with increasing $\phi_b$, \emph{i.e.,} approaching the 
transition line, thus suggesting a void structure slightly different in
the theoretical and experimental samples. 
 
\begin{figure}
{\includegraphics[width=3.2in, height=3.5in]{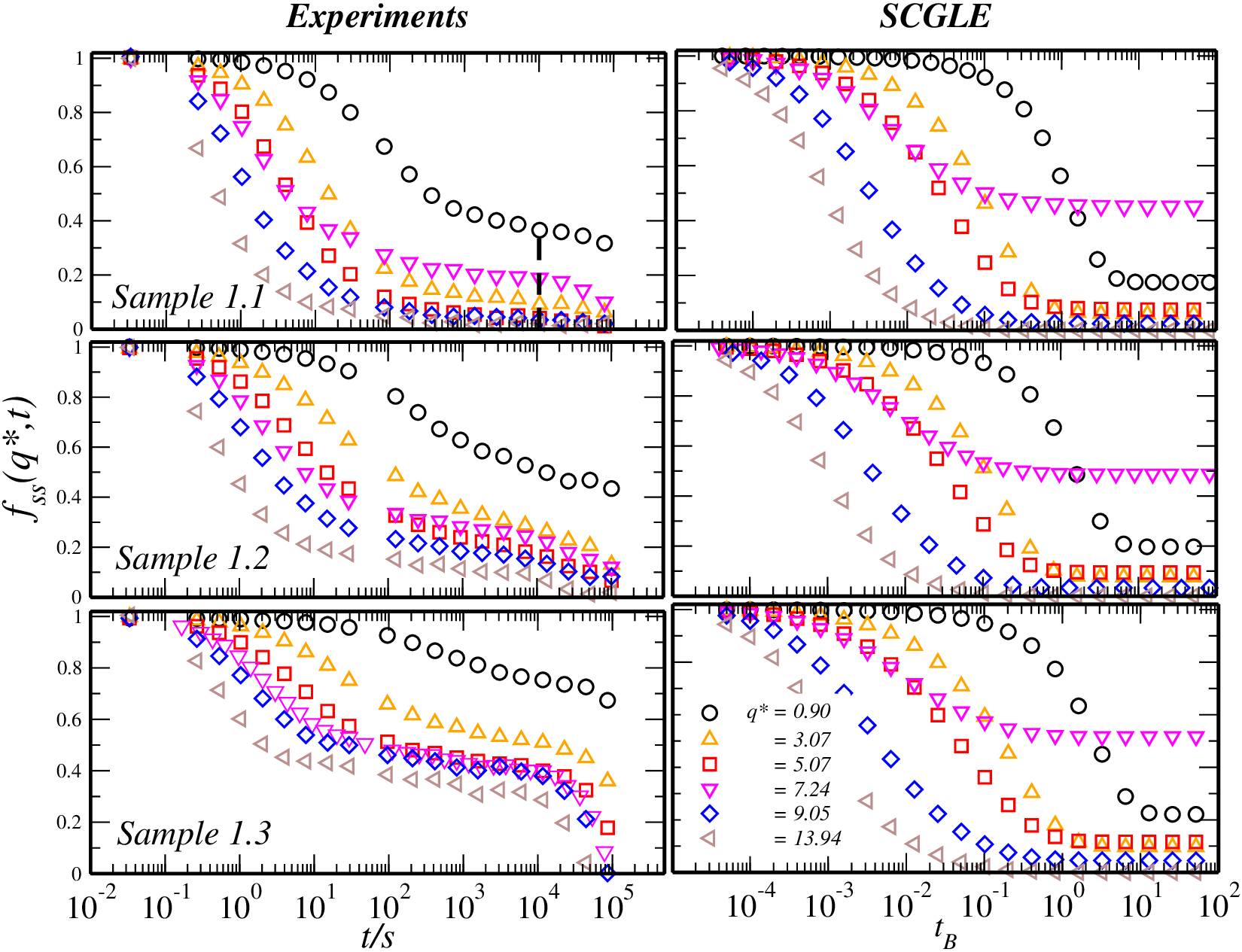}}
\caption{(Color online) Normalized collective ISFs of the small 
species, $f_{ss}(q^*,t)$, along \emph{Path 1} for different values of 
$q^{*}$. Left column: experimental results reported in Ref. 
\cite{marco3}. Right column: SCGLE predictions. The vertical dashed line 
in \emph{(a)} indicates the time, $t=10^4s$, at which the values 
$f_{ss}(q^*,t=10^5s)$ were extracted for the comparison shown in Fig. 
\ref{ergo}(a).}
\label{path1}
\end{figure}

To further investigate the influence of both $\phi_s$ and $\phi_b$ on the
height of the intermediate plateaus of $f_{ss}(q^*,t)$, we also consider
\emph{Path 1}, which corresponds to experimental data
previously reported for $\phi_s\approx  0.006$ \cite{marco3} and,  
according to the theory, located inside the \textbf{\emph{SG}} domain. 
The results displayed in Fig. \ref{path1} show that, 
in comparison to the previous case ($\phi_s \approx 0.03$), the height 
of the plateaus in $f_{ss}$ are generally shorter at comparable 
$q^*$ values; a feature that was already outlined by the SCGLE
theory in Fig. \ref{ergo}\emph{(c)}. This implies that due to crowding, 
\emph{i.e.,} increasing $\phi_s$, localization involves a larger fraction 
of small particles down to smaller length scales.

In summary, approaching the \textbf{\emph{F-SG}} transition 
($\phi_s<\phi_s^E$), the \emph{self} dynamics of the small and large 
particles decouples, with the large particles approaching dynamical 
arrest. On the other hand, the collective ISFs of both species are 
coupled at long times, with the small spheres following the arrest of 
the larger ones. In addition, their wave number dependence is modulated 
by the structure of the large particles.

\section{Dynamics from the fluid to the double glass}

We now discuss the dynamical features of the
\textbf{\emph{F}}-\textbf{\emph{DG}} transition. Theory 
and simulation results for constant $\phi_s=0.2$ and 
increasing $\phi_b$ from the fluid region towards the 
\textbf{\emph{F}}-\textbf{\emph{DG}} transition line (\emph{Path B})
are discussed. These results are complimented with results from DDM 
experiments (\emph{Path 3}).

\subsection{Self dynamics of the \emph{F-DG} transition}

\begin{figure}
{\includegraphics[width=3.2in, height=3.2in]{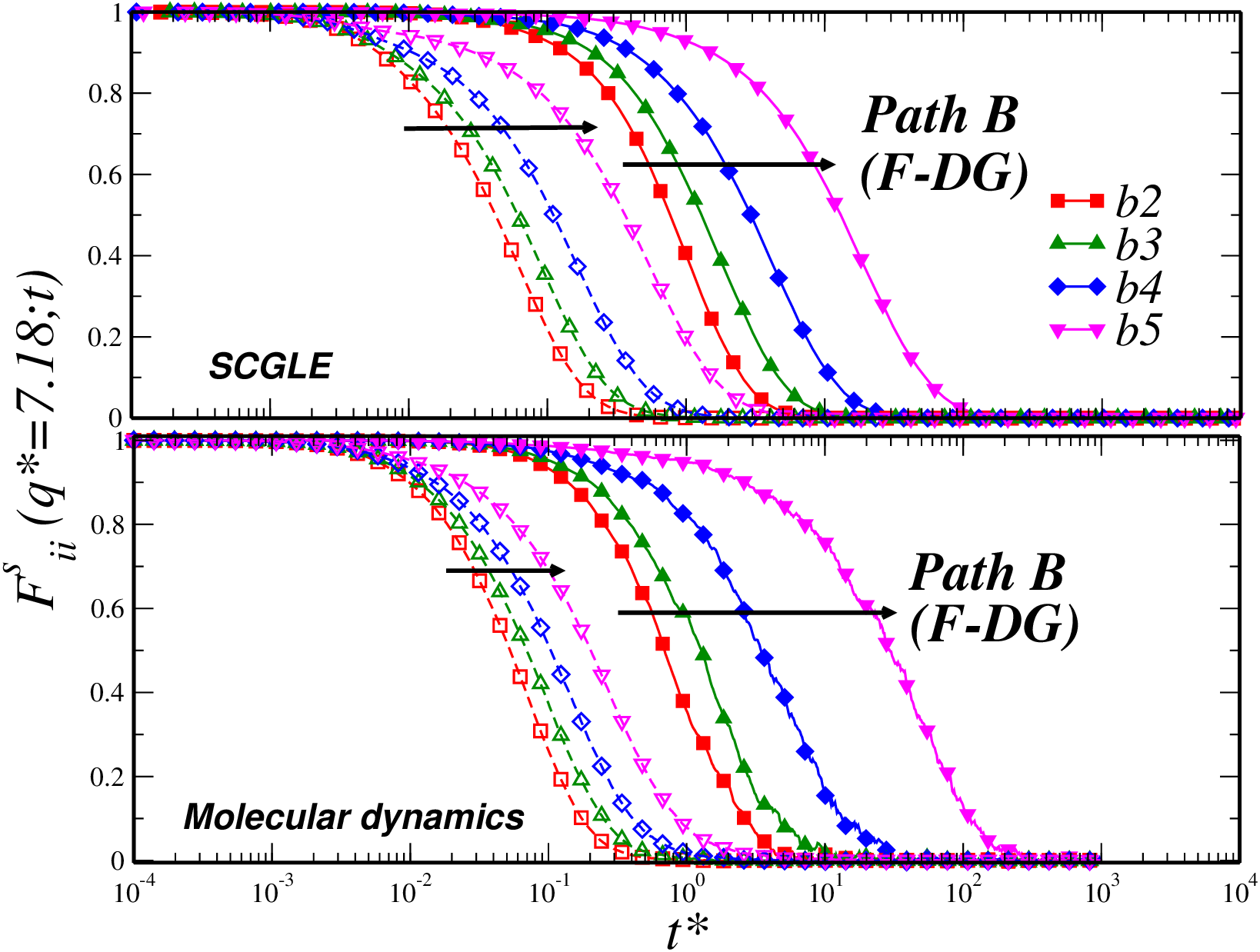}}
\caption{(Color online) Self ISFs for the large, 
$F^{S}_{bb}(q^{*},t^{*})$ (solid symbols), and small particles, 
$F^{S}_{ss}(q^{*},t^{*})$ (open symbols), calculated along 
\emph{Path B} (as indicated) at fixed $q^*=7.18$ and as 
a function of the reduced time $t^*$. The upper panel displays the 
results of the SCGLE theory
and the lower panel shows results of MD simulations.}
\label{fselfB}
\end{figure}

Fig. \ref{fselfB} reports \emph{self} ISFs for fixed $q^*=7.18$ and 
along \emph{Path B} as obtained by theory and simulations. Different
from the case of approaching to the \textbf{\emph{F}}-\textbf{\emph{SG}} 
transition (Fig. \ref{fselfA}), the shape of the \emph{self} ISFs of 
both species is rather similar, but with a faster decay for the
small particles. Upon increasing $\phi_b$, both ISFs slow down by a 
similar factor. No signature of a two-step decay in $F^S_{bb}$ or a 
long-time tail in $F^{S}_{ss}$ is observed. These results reflect the 
\emph{lubricating} effect of the large enough fraction of small particles 
on the big particles (compare, for instance, the behavior 
of $F^{S}_{bb}$ at the two state points $a5$ (Fig. \ref{fselfA}) and 
$b5$, both of which satisfy $\phi_s+\phi_b=0.65$), as suggested by 
previous experimental studies
\cite{vanmegen,vanmegen2,weeks1,weeks2,weeks3,marco1,marco2}.

\begin{figure}
{\includegraphics[width=3.2in, height=3.2in]{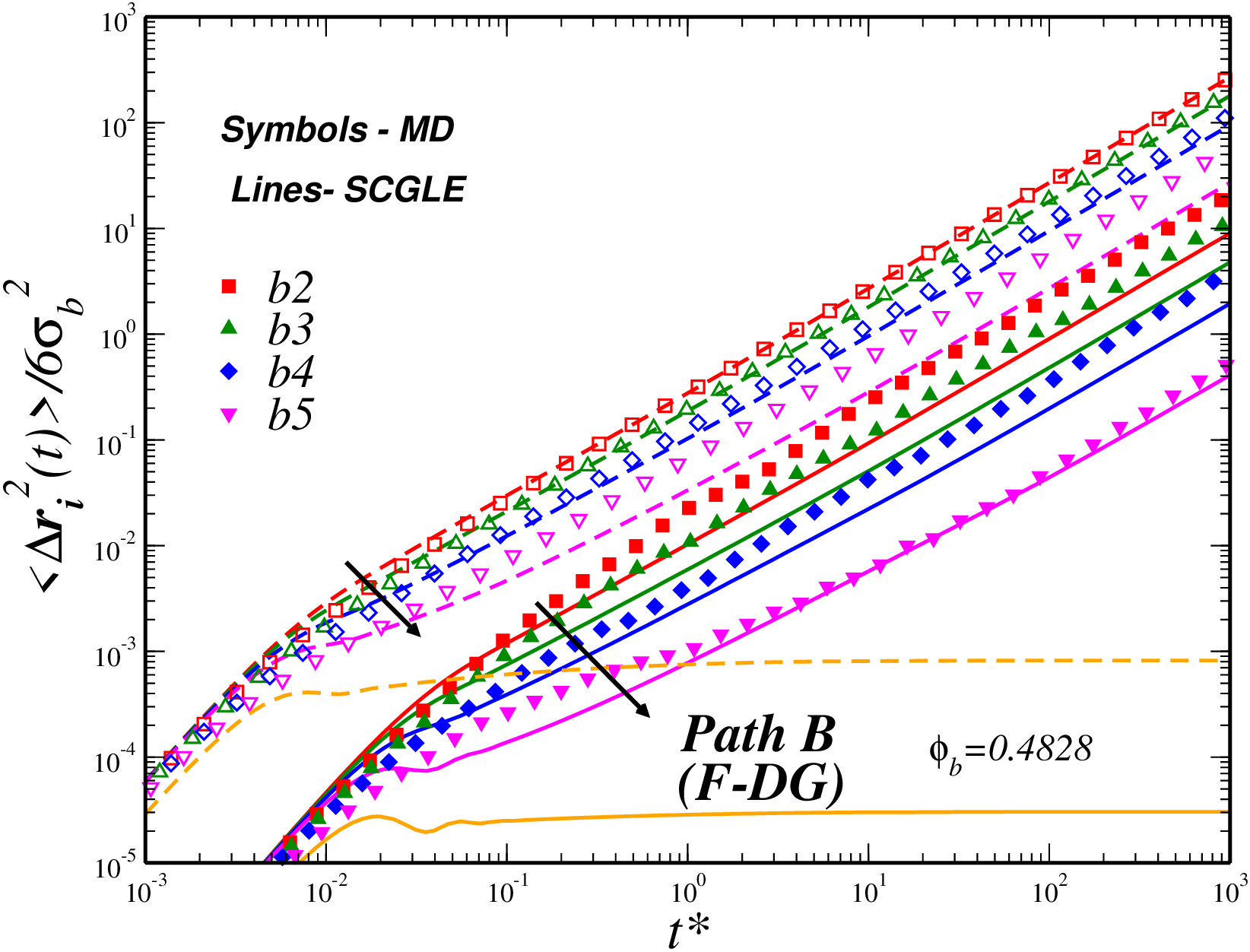}}
\caption{(Color online) MSDs for the large (solid symbols
and solid lines) and small particles (open symbols and dashed lines)
along \emph{Path B} (as indicated) obtained from MD simulations (symbols) 
and the SCGLE theory (lines).}
\label{MSDsB}
\end{figure}

These results are in agreement with the corresponding MSDs shown 
in Fig.  \ref{MSDsB}. Again, along \emph{Path B} the slow down of the 
small and large particles is very similar, in contrast to the very 
different slow down found along \emph{Path A} .

\subsection{Collective dynamics of the \emph{F-DG} transition}

\begin{figure}
{\includegraphics[width=3.2in, height=3.2in]{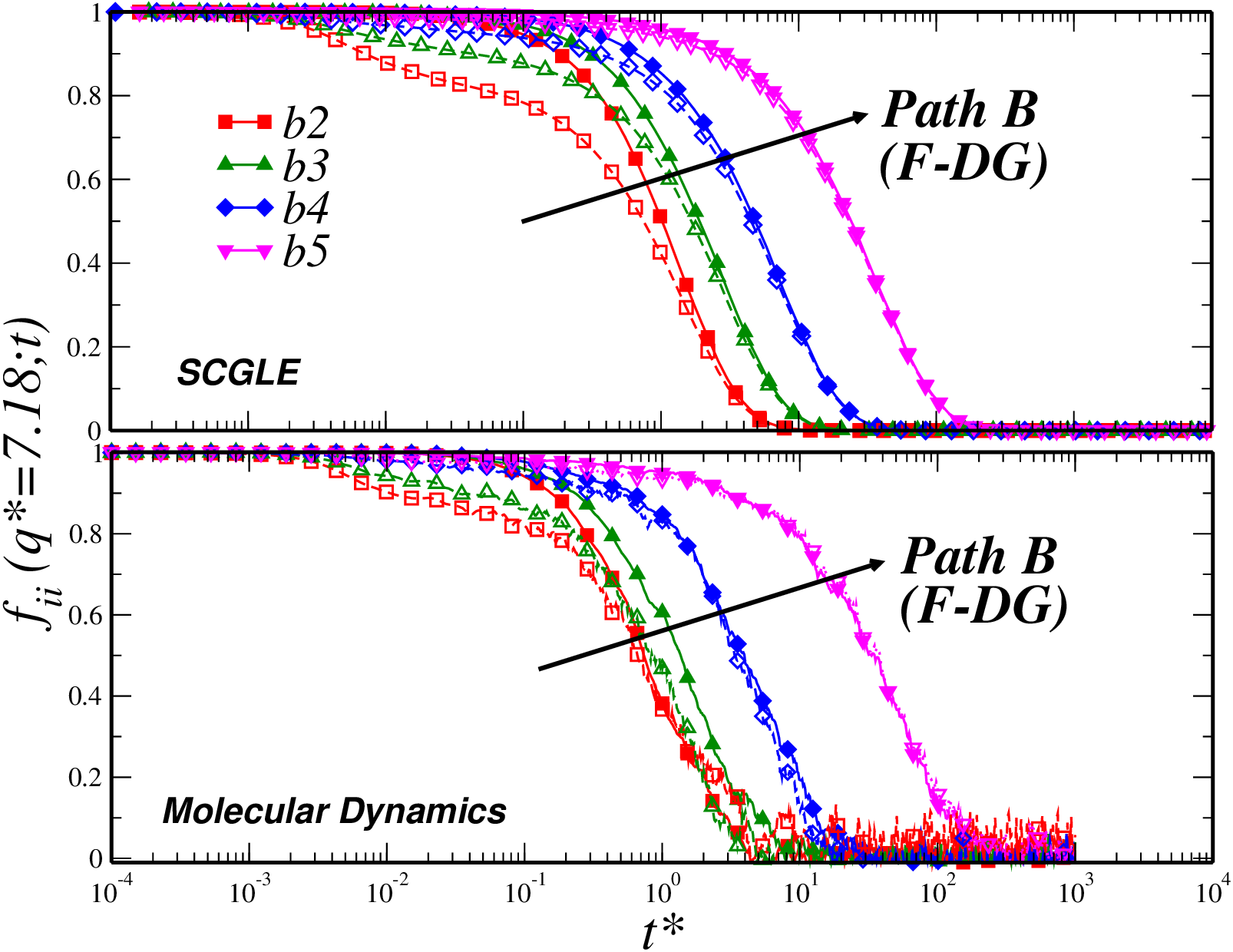}}
\caption{(Color online) Normalized collective ISFs for the large, 
$f_{bb}(q^{*},t^{*})$ (solid symbols), and small particles, $f_{ss}(q^{*},t^{*})$ (open 
symbols), calculated along \emph{Path B} (as indicated) at fixed 
$q^*=7.18$ as a function of the reduced time $t^*$. The upper panel 
displays the results from the SCGLE theory and the lower panel from MD 
simulations.} \label{fcoleB}
\end{figure}

The collective dynamics obtained by simulations and theory are displayed
in Fig. \ref{fcoleB}. 
For the large particles, $f_{bb}(q^*=7.18;t^*)$ behaves quite similar 
to the \emph{self} part, $F^{S}_{bb}(q^*=7.18;t^*)$, and also
decays faster in comparison to the behavior along \emph{Path A}. In 
contrast, for the small particles, $f_{ss}(q^*=7.18;t^{*})$ is different 
to $F^{S}_{ss}(q^*=7.18;t^*)$, but also to the behavior found along 
\emph{Path A}. The collective ISF of the small species now
show a relaxation pattern that, from intermediate times onwards, 
resembles that of the large particles $f_{bb}$. Approaching the 
\textbf{\emph{F-DG}} transition, these correlation functions become 
increasingly similar and eventually are essentially indistinguishable.

\begin{figure}
{\includegraphics[width=3.2in, height=3.2in]{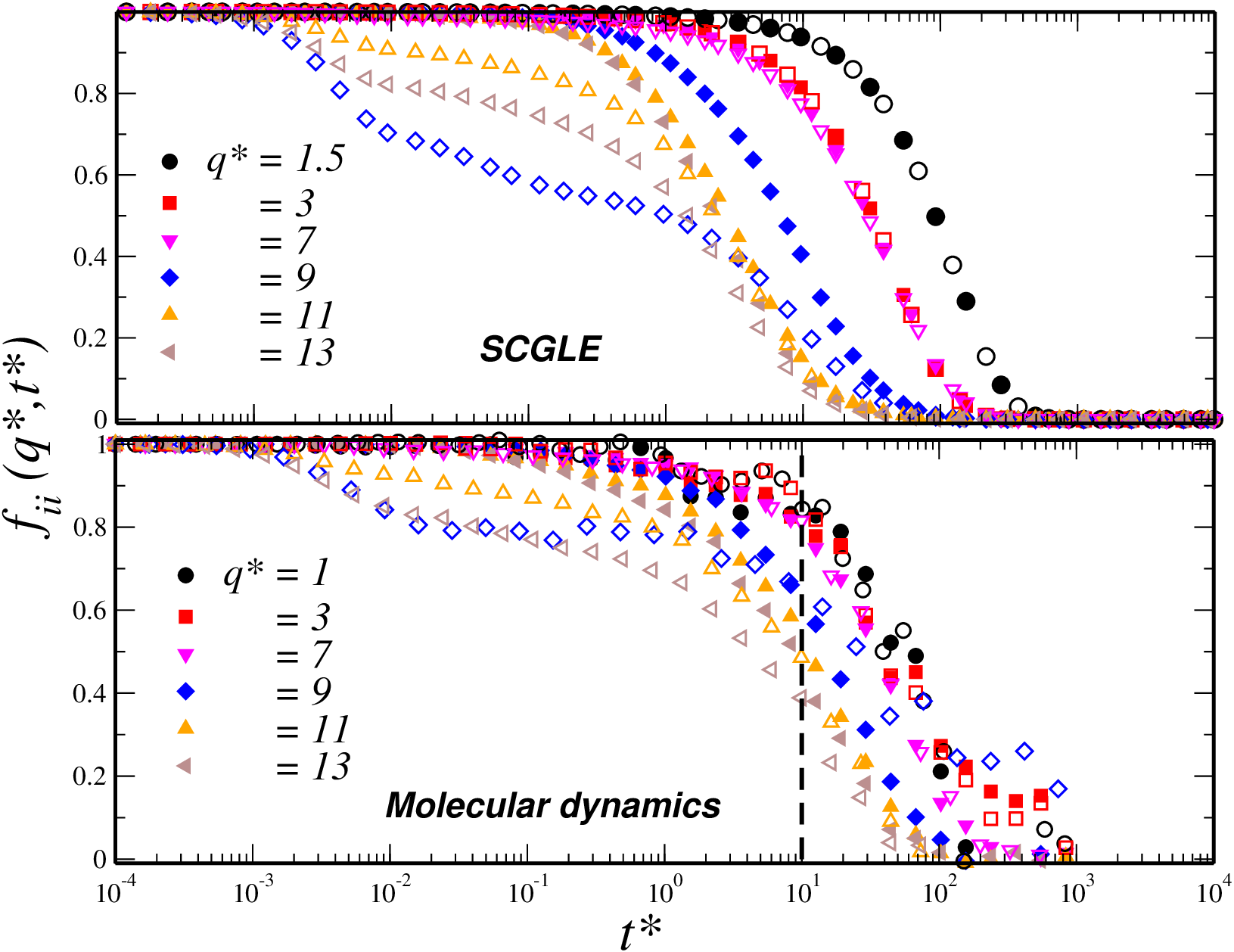}}
\caption{(Color online) Normalized collective ISFs for the large, 
$f_{bb}(q^*,t^*)$ (solid symbols), and small particles, 
$f_{ss}(q^*,t^*)$ (open symbols), at the state point 
$b5=(\phi_b=0.45,\phi_s=0.2)$, obtained from
the SCGLE theory (upper panel) and MD (lower panel). The vertical dashed
line in the lower panel indicates $t^*=10$ at which the values 
$f_{ii}(q^*,t^*=10)$ were extracted for the comparison in
Fig. \ref{ergo}(b).} \label{qdependence2}
\end{figure}

We also consider the $q^*$-dependence of the most concentrated sample,
$b5$ (Fig.\ref{qdependence2}). For $q^*\leq 7$, both $f_{bb}$ and 
$f_{ss}$ decay rather similar, with a small \emph{acceleration} of the 
dynamics of both species with increasing $q^*$.
For $q^*>7$, the relaxation time of both functions continues  
monotonically decreasing, which implies weak structural effects
on the collective dynamics, as already suggested by the results in 
Fig. \ref{ergo}(b). Inflection points, followed by intermediate plateaus, 
are observed again in $f_{ss}$. The height of these plateaus, 
however, is significantly larger than those observed for the state point 
$a5$ (Fig.\ref{qdependence}). 

\begin{figure}
\center
{\includegraphics[width=3.2in, height=3.2in]{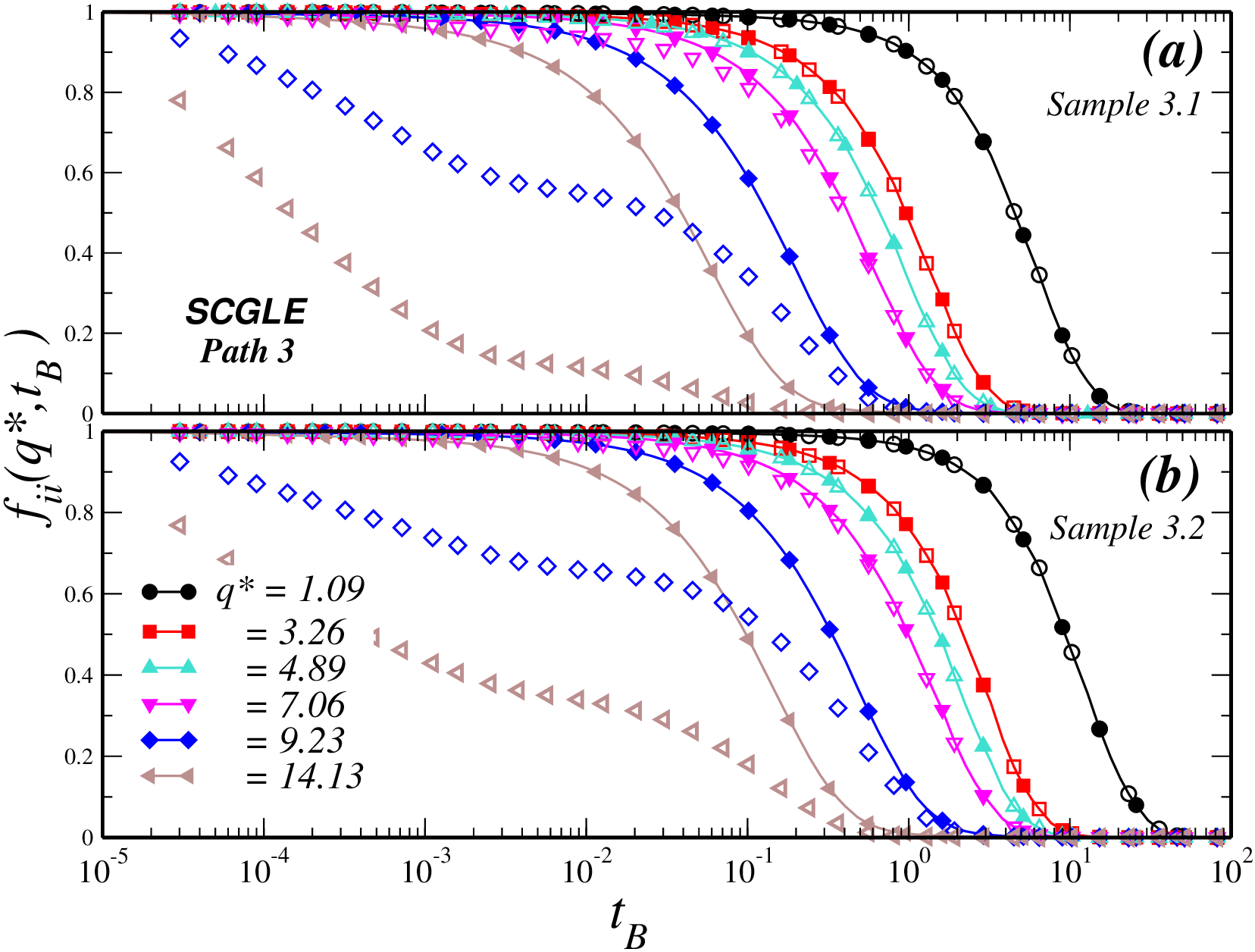}}
\caption{(Color online) Normalized collective ISFs for (a) large, $f_{bb}(q^*,t_B)$, and (b) small particles, $f_{ss}(q^*,t_B)$, along \emph{Path 3} for different reduced wave numbers $q^*$ and fixed composition (\textit{Sample 3.2}) as predicted by the SCGLE theory.} 
\label{path3_theory}
\end{figure}

To test the influence of $\phi_s$ on the $q^*$-dependence of the 
collective dynamics, we finally consider another route to the 
\textbf{\emph{F-DG}} transition, in which $\phi_b$ is kept constant 
(at about $0.3$) and $\phi_s$ is increased; \emph{i.e., Path 3}.  
The $f_{bb}(q^*,t^*)$ obtained from theory (Fig. \ref{path3_theory})
indicates a fluid behavior, with the relaxation becoming 
monotonously faster with increasing $q^*$, as observed at the state point 
$b5$. A similar situation is found for $f_{ss} (q^*,t^*)$, except for the 
distinct decay patterns observed for $q^*\geq9.23$. 

\begin{figure}
\center
{\includegraphics[width=3.2in, height=3.2in]{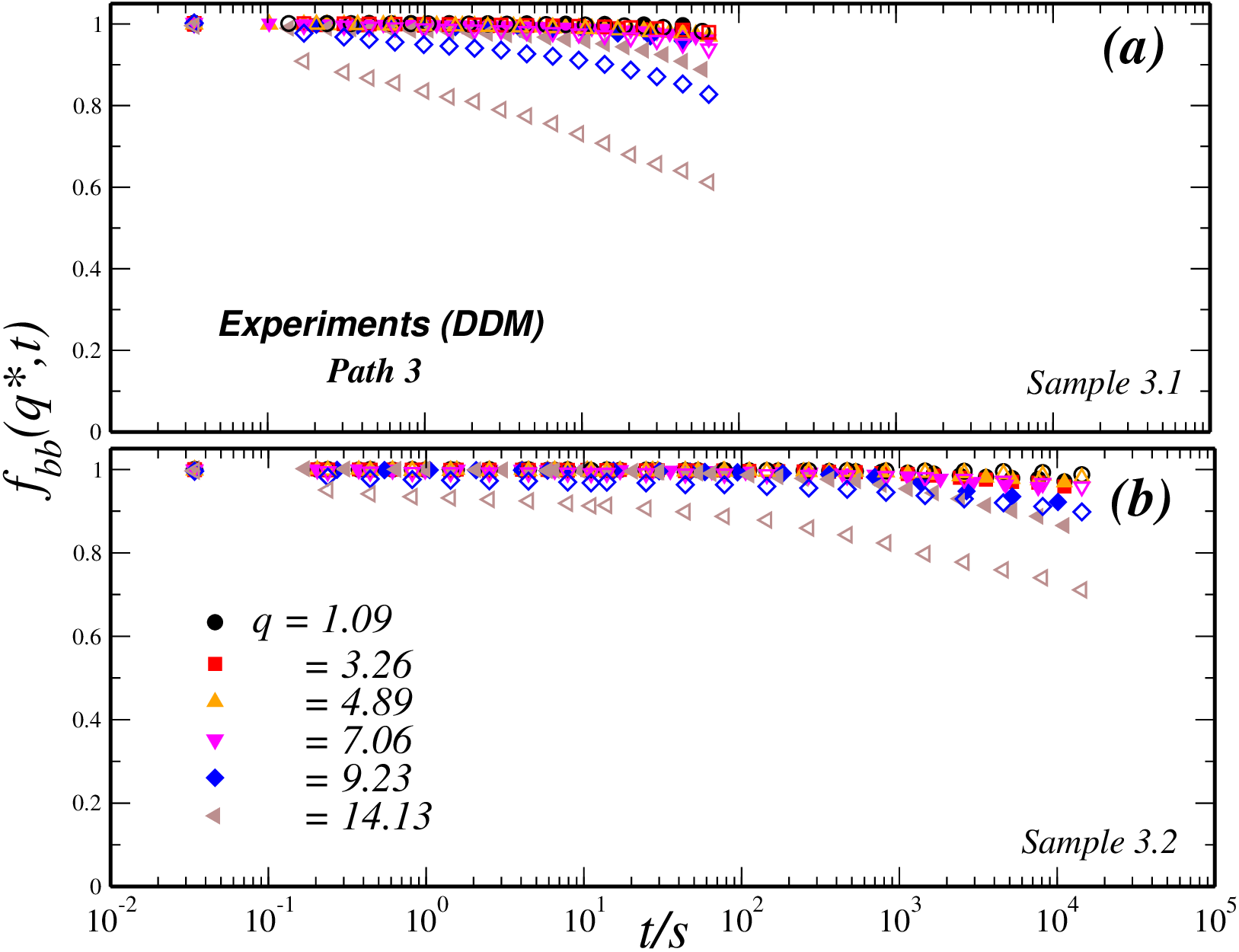}}
\caption{(Color online) Normalized collective ISFs for (a) large, $f_{bb}(q^*,t)$, and (b) small particles, $f_{ss}(q^*,t)$, along \emph{Path 3} for different reduced wave numbers $q^*$ and fixed composition (\textit{Sample 3.2}) as determined in experiments.} 
\label{path3_exp}
\end{figure}

The SCGLE results are compared with experimental data (Fig. 
\ref{path3_exp}). Due to the limited measurement time (on the order of 
a day), the experimental data extend over a smaller time-window. 
Thus, the final decays of both $f_{bb}(q,t)$ and $f_{ss}(q,t)$ are not 
accessible in the experiments. 
Nevertheless, the trends are compatible with predictions by SCGLE. 
The initial relaxation of $f_{bb}$ and $f_{ss}$ is rather similar for 
$q^*\leq7.06$, but with the later decaying faster for larger $q^*$, 
similar to the behavior found in both simulations 
(Fig. \ref{qdependence2}) and theory (Fig. \ref{path3_theory}). 
Approaching the \textbf{\emph{F}}-\textbf{\emph{DG}} transition line
(\emph{i.e.,} increasing $\phi_s$), the decay of both $f_{ss}$ and
$f_{bb}$ becomes noticeable slower at all length scales, in
contrast with the behavior towards the \textbf{\emph{F}}-
\textbf{\emph{SG}} transition (compare, for instance, the results of the
lower panel of Fig. \ref{path3_exp} against those of Figs.
\ref{path2_exp}\emph{(b)} and \ref{path2_exp}\emph(c), all of which
correspond to experimental \emph{samples} where
$\phi_b+\phi_s\approx0.6$, but with significantly different 
compositions).

In summary, the results of this section show a different scenario 
towards the \textbf{\emph{F}}-\textbf{\emph{DG}} transition, where the  
\emph{self} dynamics of the two species appear coupled and become
slower simultaneously. The collective dynamics also appears coupled 
at all the relevant length scales, but display weak structural 
effects. These results indicate that both big and small spheres become 
arrested in the \emph{self} and collective dynamics towards the 
\textbf{\emph{DG}} domain.

\section{Comparison of the dynamics towards the \emph{F-SG} and \emph{F-DG} transitions}

The self and collective dynamics approaching the \textbf{\emph{F-SG}} 
and \textbf{\emph{F-DG}} transitions have been described above and now 
will be compared. For this, notice that for each state point along 
\emph{Path B}, there is a corresponding point along \emph{Path A} having 
the same total volume fraction, $\phi$, but a different composition 
$x_s\equiv\phi_s/\phi$.
To highlight the corresponding samples, in Figs. \ref{fselfA}-
\ref{fcoleA} (\emph{Path A}) and \ref{fselfB}-\ref{fcoleB}
(\emph{Path B}) we have used the same symbols and colors for those
state points with the same $\phi$.

\begin{figure}
{\includegraphics[width=3.2in, height=3.2in]{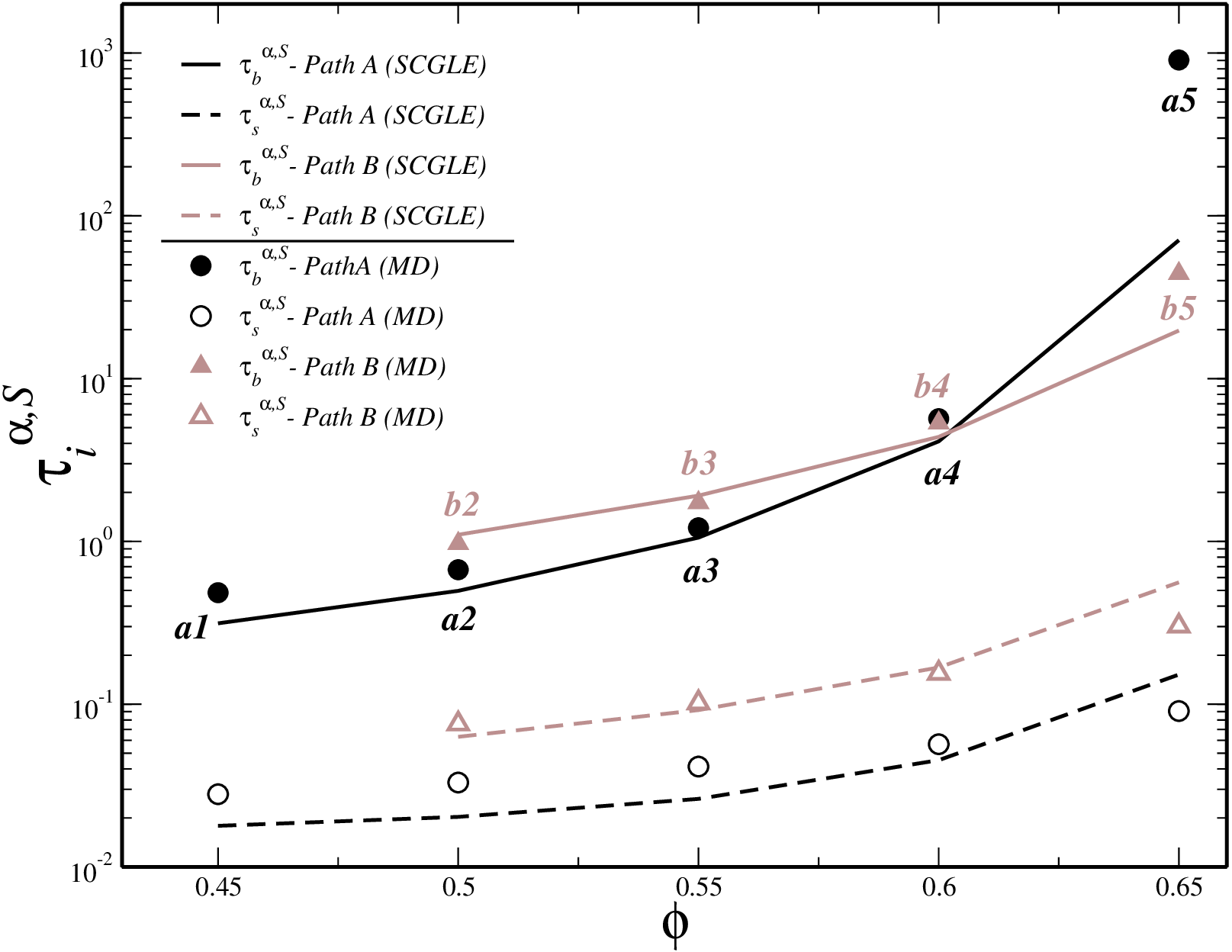}}
\caption{(Color online) \emph{Self} $\alpha$-relaxation times
of large (solid symbols, solid lines) and small particles (open symbols, 
dashed lines) along \emph{Paths A} and \emph{B} as a function 
of the total volume fraction $\phi$. 
Symbols represent MD data and lines SCGLE results, as indicated.} 
\label{taus}
\end{figure}

In Fig. \ref{taus}, the evolution of the \emph{self} dynamics along 
\emph{Paths A} and \emph{B} is compared in terms of the 
$\alpha$-relaxation times $\tau^{\alpha,S}_i$, defined here as 
$F^S_{ii}(q^*=7.18,\tau^{\alpha,S}_i)=1/e$. 
For samples with the same $\phi$, the relaxation of the small species 
is moderately slower along \emph{Path B} ($\phi_s=0.2$) than 
\emph{Path A} ($\phi_s=0.05$), but displaying essentially the same 
dependence on $\phi$. This indicates weak effects of the composition 
on the \emph{self} dynamics of the small particles towards the glass 
transition, despite the different relaxation patterns observed in 
$F^S_{ss}$ along \emph{Paths A} (Fig.\ref{fselfA}) and \emph{B} 
(Fig. \ref{fselfB}).
A distinct behavior for the large species is found. For $\phi<0.6$, the 
difference in the relaxation along both paths is smaller in comparison to 
the case of the small species, with the decay being again slightly slower 
along \emph{Path B}. A crossover, however, is observed at 
$\phi\approx0.6$ and, for larger $\phi(=0.65)$, the relaxation becomes 
noticeable slower along \emph{Path A}, which suggest strong effects of 
both composition and total concentration for the large particles' 
\emph{self} dynamics.

\begin{figure}
{\includegraphics[width=3.2in, height=3.2in]{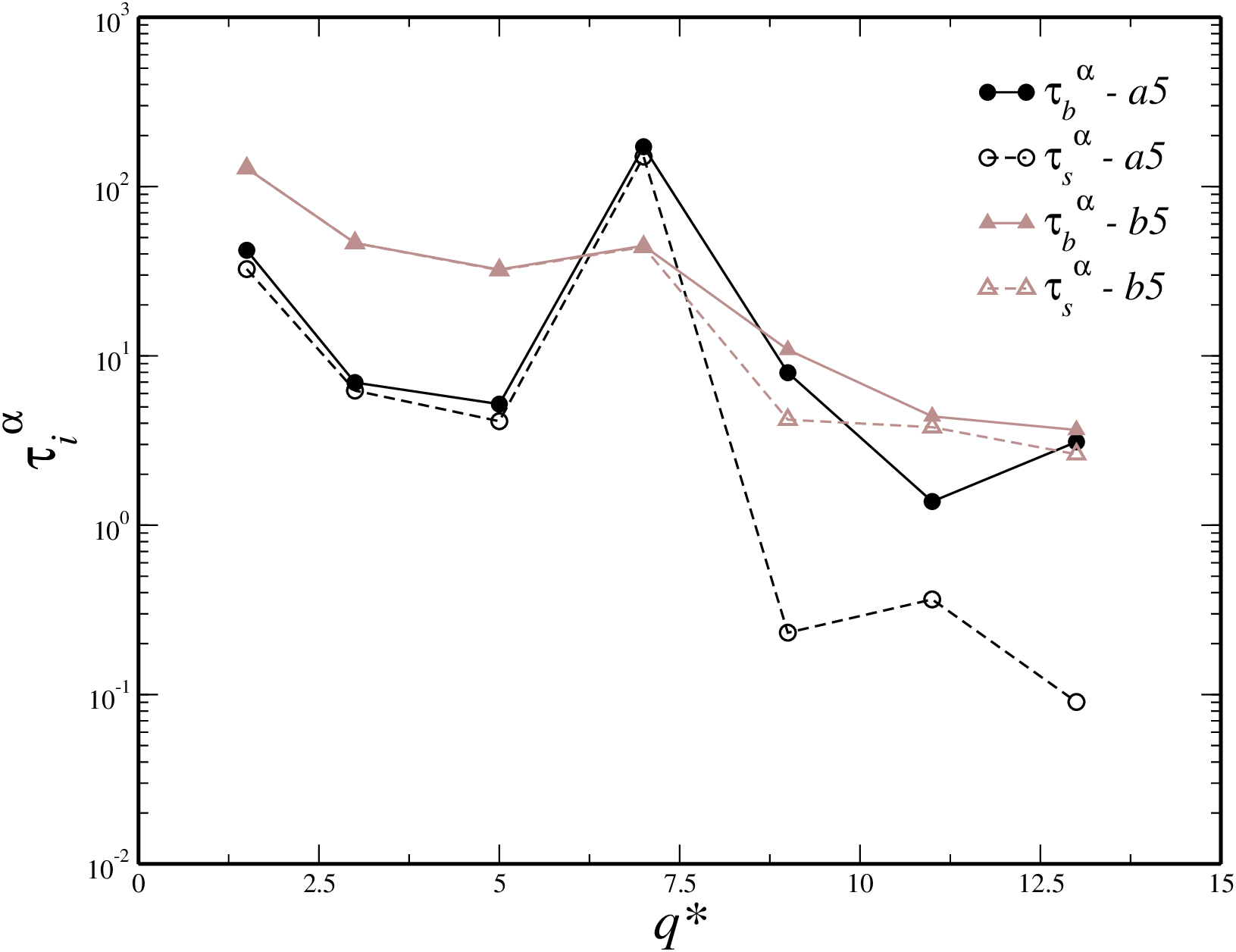}}
\caption{(Color online) Collective $\alpha$-relaxation times of both 
species as a function of reduced wave number $q^*$ at the state points 
$a5$($\CIRCLE$-$\tau_b^{\alpha}$, $\Circle$-$\tau_s^{\alpha}$) and $b5$
(\textcolor{antiquebrass}{$\blacktriangle$}-$\tau_b^{\alpha}$,
 \textcolor{antiquebrass}{$\bigtriangleup$}-$\tau_s^{\alpha}$), as 
predicted by the SCGLE approximation.}
\label{tauscole}
\end{figure}

To compare the collective dynamics close to the 
\textbf{\emph{F}}-\textbf{\emph{SG}} and 
\textbf{\emph{F}}-\textbf{\emph{DG}} transitions, we similarly consider 
collective relaxation times defined by 
$f_{ii}(q^*,t^*=\tau^{\alpha}_i)=1/e$, for different $q^*$-values, and 
at the state points $a5$ and $b5$, respectively (Fig. \ref{tauscole}). 
For $q^*\leq7.18$, the relaxation times of the small and large particles 
are strongly coupled for both compositions, with \emph{Path B} 
displaying the slower relaxation for $q^*<7$ and a crossover at 
$q\approx7.18$ similar to the behavior of $\tau_i^{\alpha,S}$.
At $q^*>7.18$ and small $x_s$ \emph{i.e., many big particles} the 
relaxation of each species appears decoupled. The small species decay
faster, indicating their ability to explore the local environment.
If the fraction of the small particles is increased, the relaxation 
remains coupled, which reflects a more pronounced contribution of 
the small particles' dynamics to the slowing down.

\section{Concluding remarks}\label{conclusions}

By combining experiments, molecular dynamics simulations and theoretical 
calculations based on the SCGLE formalism, we have 
built up a general and consistent physical description of glassy 
dynamics in highly asymmetric binary mixtures of hard-spheres. Two 
fundamentally different scenarios for the glass 
transition were confirmed. Both lead to dynamical arrest; they are 
related to distinct microscopic states that, however, 
also bear similar qualitative features.
 
Below a certain volume fraction of small spheres, $\phi^E_s$, a single 
glass occurs, where the slow dynamics is dominated by the large species 
and whose \emph{self} and collective dynamics indicate 
structural arrest with the typical fingerprints of an ideal glass 
transition of hard spheres. In contrast, the \emph{self} dynamics of 
the small species does not arrest, but resembles the behavior of a fluid 
embedded in a heterogeneous  medium, including 
the long-time tail in the relaxation. Their collective 
dynamics at large length scales appeared coupled to the big particles and 
the dependence on the wave number follows the structure factor of the 
large particles, whereas at short length scales the dynamics remained 
ergodic. This is consistent with the following observation: the 
total volume fraction to reach the glass transition becomes larger 
with increasing the volume fraction of small particles, $\phi_s$. 

In the other possible glass transition referred to as a double glass, 
the slowing down of the \emph{self} and collective dynamics of both 
species appeared strongly correlated. In the latter case, the coupling  
occurred at all length scales, i.e., both species become dynamically 
arrested simultaneously.
Contrary to the previous case, the total volume fraction at which 
the double glass transition occurs became smaller with 
increasing $\phi_s$.

These fundamental differences when the mixture approaches to any glass 
boundary strongly suggest that the degrees of freedom of the small 
particles cannot be simply integrated out. For instance, assuming an 
effective potential between the large particles neglects the dynamical 
contribution of the small particles.

The above features were stressed out by all the techniques here 
employed, allowing to summarize our findings in a dynamical arrest 
diagram. Then, our results might serve to locate and reinterpret previous
results in one of the possible glassy domains and as benchmark for 
future tests in colloidal hard sphere mixtures and other systems with 
competing time and length scales, such as metallic alloys, polymers or
protein solutions. Finally, from a technological point of view, the 
understanding of arrested states will facilitate rational design 
of new materials, as well as many industrial processes.  

\section*{Acknowledgments}
This work was supported by the Consejo Nacional de
Ciencia y Tecnolog\'{\i}a (CONACYT, Mexico) through
grants Nos. 242364, 182132, 237425, 358254, 
FC-2015-2/1155, LANIMFE-279887-2017, and CB-2015-01-257636.
L.F.E.A acknowledges financial support from the
German Academic Exchange Service (DAAD) through the DLR-DAAD
programme under grant No. 212. P. L. and S. U. E. acknowledge 
support by the German-Israeli Foundation 
(Grant No. J-1345-303.10/2016). R. C. P. acknowledges the financial 
support provided by Marcos Moshinsky Foundation, the University of 
Guanajuato (Convocatoria Institucional de Investigaci\'on Cient\'ifica 
2018), and the Alexander von Humboldt Foundation during his stay at the
University of D\"usseldorf in summer 2018.


\begin{thebibliography}{99}

\bibitem{greer} A. L. Greer, \emph{Science}, Vol. \textbf{267}, (1995) p.
1947-1953.

\bibitem{zahid} S. Zahid \emph{et.al.,}  RSC Adv., 2016, \textbf{6},
70197-70214

\bibitem{takahashi} Y. Takahashi, Y. Yamazaki. R. Ihara and T. Fujiwara,
\emph{Sci. Rep.}
\textbf{3}, 1147; DOI:10.1038/srep01147 (2013).

\bibitem{hodge} I. M. Hodge, \emph{Science}, Vol. \textbf{267}, (1995) p.
1945-1947.

\bibitem{cipelletti1} L. Cipelletti and L. Ramos, \emph{J. Phys.: Condens.
Matter} \textbf{17},  R253 (2005).

\bibitem{weitz}P.N. Segr\'{e}, V. Prasad, A. B. Schofield, and D. A.
Weitz,
\emph{Phys. Rev. Lett.}, \textbf{86}, No. 26, 6042 (2001).

\bibitem{lu} P. J. Lu, E. Zacarelli, F. Ciulla, A. B. Schofield, F.
Sciortino and D. A. Weitz,
\emph{Nature}, 2008, 453 (7194) 499-503, doi: 10.1038/nature06931.

\bibitem{zacarelli1} E. Zacarelli and W.C. K. Poon (2009),
\emph{Proc. Natl. Acad. Sci. USA}, 106: 15203-15208.

\bibitem{norman} A.P.R. Eberle, N.J. Wagner and R. Casta\~neda-Priego, 
\emph{Phys. Rev. Lett.,} \textbf{106}, 105704 (2011).

\bibitem{angell} C. A. Angell, \emph{Science}, Vol. \textbf{267}, (1995)
p. 1924-1935.

\bibitem{angell1} C. A. Angell, K. L. Ngai , G. B. McKenna, P. F. 
McMillan and S. F Martin., \emph{J. Appl. Phys.}
\textbf{88}, 3113 (2000).

\bibitem{weeks1} J. M. Lynch, G.C. Cianci and E. R. Weeks, \emph{Phys.
Rev. E}, \textbf{78}, 031410 (2008).

\bibitem{weeks2} T. Narumi, S.V. Franklin, K.W. Desmond, M. Tokuyama and
E.R. Weeks,
\emph{Soft Matter}, 2011, \textbf{7}, 1472-1482.

\bibitem{weeks3} S. Vivek, C.P. Kelleher, P.M. Chaikin and E.R. Weeks,
\emph{Proc. Natl. Acad. Sci.
USA}, Vol. \textbf{114}, no.8, 1850-1855 (2017).

\bibitem{vanmegen1} W. van Megen and P.N. Pusey, \emph{Phys. Rev. A},
\textbf{43}, No. 10, 5429, (1991).

\bibitem{vanmegen2} W. van Megen, S.M. Underwood and P.N. Pusey,
\emph{Phys. Rev. Lett.}
\textbf{67}, No.12, 1587 (1991).

\bibitem{sciortino} F. Sciortino and P. Tartaglia, \emph{Adv. Phys.}
\textbf{54}, 471 (2005).

\bibitem{hunter} G.L. Hunter and E.R. Weeks, \emph{Rep. Prog. Phys.}
\textbf{75} (2012) 066501.

\bibitem{jan} A. Imhof and J. K. G Dhont, \emph{Phys. Rev. Lett.},
\textbf{75}, No.8, 1662 (1997).

\bibitem{vanmegen}S.R. Williams and W. van Megen, \emph{Phys. Rev. E}
\textbf{64}, 041502 (2001).

\bibitem{pham} K.N. Pham, \textit{et al., Science}, Vol. \textbf{296}, 104
(2002).

\bibitem{pham1} K. N. Pham, S. U. Egelhaaf, P. N. Pusey, and W. C. K.
Poon, \emph{Phys. Rev. E} \textbf{69}, 011503 (2004).

\bibitem{eckert} T. Eckert and E. Bartsch, \emph{Phys. Rev. Lett.}
\textbf{89}, 125701 (2002).

\bibitem{marco1} T. Sentjabrskaja, M. Hermes, W. C. K. Poon, C. D.
Estrada, R. Casta\~neda-Priego, S.U. Egelhaaf and M. Laurati,
\emph{Soft Matter}, 2014, \textbf{10}, 6546.

\bibitem{marco2} J. Hendricks, R. Capellmann, A. B. Schofield, S. U.
Egelhaaf, and M. Laurati, \emph{Phys. Rev. E} \textbf{91}, 032308 (2015).

\bibitem{ameyer} A. Meyer \emph{Phys. Rev. B}, \textbf{66} 13425, (2002).

\bibitem{thvg1} Th. Voigtmann, A. Meyer, D. Holland-Moritz, S. St\"uber, 
T. Hansen and T. Unruh, 2008 \emph{EuroPhys. Lett.} \emph{82} 66001

\bibitem{weeks} E. R. Weeks and D. A. Weitz, \emph{Phys. Rev. Lett.,} 
Vol. \textbf{89} 095704 (2002).

\bibitem{marco3} T. Sentjabrskaja, \emph{et al.,} Nat. Comm. 7:11133 doi:
10.1038/ncomms11133 (2016).

\bibitem{marco4} D. Heckendorf, K.J. Mutch, S.U. Egelhaaf and M. Laurati,
\emph{Phys. Rev. Lett.}, \textbf{119}, 048003 (2017).

\bibitem{moreno} A.J. Moreno and J. Colmenero, \emph{J. of Chem Phys} 
\textbf{125}, 164507 (2006).

\bibitem{mayer} C. Mayer, F. Sciortino, C.N. Likos, P. Tartaglia,
Hartmut L\"owen and E. Zacarelli, \emph{Macromolecules}, 2009, 42(1),
pp 423-434.

\bibitem{horbach} Th. Voigtmann and J. Horbach, \emph{Phys. Rev. Lett.} 
\textbf{103}, 205901 (2009).

\bibitem{fabian} L. Fabbian. W. G\"otze, F. Sciortino, P. Tartaglia, and
F. Thiery, Phys. Rev. E, \textbf{59}, No.2 R1347 (1999).

\bibitem{fuchs} J. Bergenholtz and M. Fuchs, \emph{Phys. Rev. E,} 
\textbf{59}, No.5, 5709, (2009).

\bibitem{dawson} K.A. Dawson, \emph{Curr. Opin. Colloid Interf. Sci.}
\textbf{7}, 218 (2002).

\bibitem{nagi1} N. Khali, A. de Candia, A. Fierro, M. Pica Ciamarra and 
A. Coniglio, \textit{Soft Matter}, 2014, \textbf{10}, 4800.

\bibitem{Erik2013} E. L\'opez-S\'anchez, C. D. Estrada-Alvarez, G.
P\'erez-Angel, J. M. M\'endez-Alcaraz, P. Gonz\'alez-Mozuelos and
R. Casta\~neda-Priego, \emph{J. Chem. Phys.} \textbf{139}, 104908 (2013).

\bibitem{dijkstra} C. Grodon, M. Dijkstra, R. Evans, R. Roth, \emph{J.
Chem. Phys.}, \textbf{121}, 7869 (2004).

\bibitem{goetze1}  W. G\"{o}tze, in {\em Liquids, Freezing
and Glass Transition}, edited by J. P. Hansen, D. Levesque, and J.
Zinn-Justin (North-Holland, Amsterdam, 1991).

\bibitem{goetze2} W. G\"{o}tze and L. Sj\"ogren, Rep. Prog. Phys. {\bf
55}, 241 (1992).

\bibitem{goetze3} W. G\"otze, \emph{Condened Matt. Phys}, 1998, Vol. 1,
No. 4(16), p. 873-904.

\bibitem{goetze4} W. G\"otze, \emph{Complex Dynamics of Glass-Forming
liquids. A Mode-coupling theory}, Oxford University Press (2009).

\bibitem{yeomans0} L. Yeomans-Reyna and M. Medina-Noyola, \emph{Phys. 
Rev. E}, \textbf{640}, 066114 (2001).

\bibitem{yeomans1} L. Yeomans-Reyna, \emph{et al.}, \emph{Phys. Rev. E}
\textbf{76}, 041504 (2007).

\bibitem{chavez} M. A. Ch\'avez-Rojo and M. Medina-Noyola, \emph{Phys.
Rev. E} \textbf{72}, 031107 (2005),
Phys. Rev. E \textbf{76}, 039902 (2007).

\bibitem{todos} R. Ju\'{a}rez-Maldonado {\it et al.}, \emph{Phys. Rev. E}
{\bf 76}, 062502 (2007).

\bibitem{rigo1} R. Ju\'{a}rez-Maldonado and M. Medina-Noyola,
\emph{Phys.Rev. E.}, \textbf{77}, 051503 (2008).

\bibitem{rigoAO} R. Ju\'{a}rez-Maldonado and M. Medina-Noyola,
\emph{Phys.Rev. Lett.} \textbf{101}, 267801 (2008).

\bibitem{thvggotz} Th. Voigtmann and W. G\"otze, \emph{Phys. Rev. E},
\textbf{67}, 021502 (2003).

\bibitem{thomas} Th. Voigtmann, 2011 \emph{EuroPhys. Lett.} 
\textbf{96}, 36006.

\bibitem{percus} J. K. Percus and G. J. Yevick, \emph{Phys. Rev.},
\textbf{110}, 1 (1957).

\bibitem{verlet} L. Verlet and J.-J. Weis, \emph{Phys. Rev. A},
\textbf{5}, 939 (1972).

\bibitem{porous} L.F. Elizondo-Aguilera and M. Medina-Noyola, \emph{J.
Chem. Phys.,} \textbf{142} 224901 (2015).

\bibitem{edilio} E. L\'azaro-L\'azaro, P. Mendoza-M\'endez,  L.F.
Elizondo-Aguilera, J.A. Perera-Burgos, P. E. Ram\'irez-Gonz\'alez,
G. P\'erez-\'Angel, R. Casta\~neda-Priego and
M. Medina-Noyola, \emph{J. Chem. Phys.,} \textbf{146} 184506 (2017).

\bibitem{vargas-PRE-2013} M.\ C.\ Vargas and G.\ P\'erez-\'Angel,
\emph{Phys.\ Rev.\ E.} {\bf 87}, 042313 (2013).

\bibitem{nescgle1} P. E. Ram\'irez-Gonz\'alez and M. Medina-Noyola, Phys. Rev. E \textbf{82}, 061503 (2010).

\bibitem{nescgle4}  L. E. S\'anchez-D\'iaz, E. L\'azaro-L\'azaro, J. M. Olais-Govea and M. Medina-Noyola,
J. Chem Phys.  \textbf{140}, 234501 (2014).

\bibitem{nescgle6}  P. Mendoza-M\'endez, E. L\'azaro-L\'azaro, L. E. S\'anchez-D\'iaz, P. E. Ram\'irez-Gonz\'alez,
G. P\'erez-\'Angel, and M. Medina-Noyola, Phys. Rev. E \textbf{96}, 022608 (2017).


\end{thebibliography}
\end{document}